\documentclass[useAMS,usenatbib]{mnras}
\usepackage{times}
\usepackage[normalem]{ulem}
\usepackage{amssymb}
\usepackage{graphicx}
\usepackage{amsmath}
\usepackage[usenames]{color}
\definecolor{DarkGreen}{rgb}{0.64,0.80,0.35}
\definecolor{pink}{rgb}{1,0.60,0.60}
\definecolor{lightblue}{rgb}{0.,0.6,.60}
\newcommand{\tobs}{$t_{\mathrm{obs}}$\,}

\title[Flares from long-lived reverse shocks]{ Relativistic simulations of long-lived reverse shocks in stratified ejecta: the origin of flares in GRB afterglows}

\author[A. Lamberts and F. Daigne]{A. Lamberts$^{1}$\thanks{E-mail:lamberts@caltech.edu} and F. Daigne$^{2}$ \\
$^{1}$ Theoretical Astrophysics, California Institute of Technology, Pasadena, CA 91125, USA\\
$^{2}$
UPMC-CNRS, UMR7095, Institut d'Astrophysique de Paris, F-75014 Paris, France
}
\begin{document}
\date{\today}

\pagerange{\pageref{firstpage}--\pageref{lastpage}} \pubyear{2017}

\maketitle

\label{firstpage}

\begin{abstract}
The X-ray light curves of the early afterglow phase from gamma-ray bursts present a puzzling variability, including flares.  The origin of these flares is still debated, and often associated with a late activity of the central engine. We discuss an alternative scenario where the central engine remains short-lived and flares are produced by the propagation of a long-lived reverse shock in a stratified ejecta. Here we focus on the hydrodynamics of the shock interactions. We perform one-dimensional ultrarelativistic hydrodynamic simulations with different initial internal structure in the gamma-ray burst ejecta. We use them to extract bolometric light curves and compare with a previous study based on a simplified ballistic model. We find a good agreement between both approaches, with similar slopes and variability in the light curves, but identify several weaknesses in the ballistic model: the density is underestimated in the shocked regions, and more importantly, late shock reflections are not captured. With accurate dynamics provided by our hydrodynamic simulations, we confirm that internal shocks in the ejecta lead to the formation of dense shells.  The interaction of the long-lived reverse shock with a dense shell then produces a fast and intense increase of the dissipated power. Assuming that the emission is due to the synchrotron radiation from shock-accelerated electrons, and that the external forward shock is radiatively inefficient, we find that this results in a bright flare in the X-ray lightcurve, with arrival times, shapes, and duration in agreement with the observed properties of X-ray flares in GRB afterglows.
\end{abstract}


\section{Introduction}

Gamma-ray bursts (GRB) are intense flashes of gamma-rays of extragalactic origin, 
with an apparent rate of about once a day. Their duration shows a bimodal distribution, with long bursts (from a few seconds to  a few minutes) associated with the collapse of certain massive stars \citep{1993ApJ...405..273W,1999ApJ...524..262M}, and short bursts (from a few milliseconds to a few seconds) believed to originate from mergers of compact objects in a binary system \citep[see][for reviews]{2004RvMP...76.1143P,2012ConPh..53..339G,2012Sci...337..932G,2014ARA&A..52...43B}. The prompt emission peaks in the keV-MeV range and is followed by a rapidly fading afterglow observed on longer timescales from the X-ray to the radio wavelengths \citep[see] [for a recent review]{2016SSRv..202....3Z}. 

The initial event leads to the formation of a new compact source, probably an accreting black hole, even if rapidly rotating magnetars are also discussed \citep{2007ApJ...665..599T,2013MNRAS.430.1061R}. The highly variable prompt emission corresponds to a tremendous energy radiated in gamma-rays ($E_\mathrm{iso}\sim 10^{51}-10^{54}\, \mathrm{erg}$) and must be due to internal dissipation (to account for the fast variability) in a ultra-relativistic outflow (to avoid a strong $\gamma\gamma$ annihilation).

The afterglow phase is due to the deceleration of the relativistic outflow by the external medium. In the standard scenario, the afterglow emission is due to the synchrotron radiation of non-thermal electrons accelerated at the ultra-relativistic (external) forward shock \citep[see][for a recent review]{2015PhR...561....1K}. This naturally leads to an observed flux $F_\nu\propto t_\mathrm{obs}^{-\alpha}$ with $\alpha\simeq 1$ \citep{1997ApJ...476..232M,1998ApJ...497L..17S}, in agreement with observations a few hours after the burst. 

However, this simple picture was soon challenged by the first observations with the \textit{Swift} satellite \citep{2004ApJ...611.1005G}, which revealed a great diversity of light curves, especially in the early afterglow phase, and an unexpected variability in the X-ray afterglow. The latter can show an early steep decay just after the prompt emission ($\alpha\simeq 3$), often followed by a plateau phase ($\alpha\simeq 0$) before the standard decay ($\alpha\simeq 1$) \citep{2005Natur.436..985T,2006ApJ...642..389N,2006ApJ...647.1213O}. In addition, X-ray flares are observed in $\sim 30\%$ of GRBs \citep[see e.g][]{2005SSRv..120..165B,2010MNRAS.406.2113C}, with some bursts presenting several flares between a few $10\, \mathrm{s}$ and several $10^4$\,s after the trigger. The underlying shape of the X-ray light curve remains unaffected by the flares.  The flares are asymmetric, with a steep rise and a slower decay \citep{2007ApJ...671.1903C}.  However, contrary to the gamma-ray variability observed during the prompt phase, the X-ray flares present  a common behavior with duration increasing with time, following $\Delta t_{\mathrm{obs}}/t_{\mathrm{obs}}\simeq 0.1-0.3 $. 

This observed diversity and variability of the X-ray afterglow is difficult to reconcile with the standard external shock model. In particular, the emission is only mildly sensitive to the structure of the external medium \citep[e.g.][]{2002ApJ...565L..87D,2005ApJ...631..435R,2011MNRAS.410.1064M,2013ApJ...773....2G}. Plateaus can be reproduced with late energy injection :\citep{1998ApJ...496L...1R,2000ApJ...535L..33S,2017ApJ...835..206L} and flares can be associated with late internal dissipation in the ejecta in the case of a long-lasting central engine \citep[e.g.][]{2005Sci...309.1833B,2006ApJ...642..354Z,2005MNRAS.364L..42F,2011MNRAS.410.1064M}. Such possibilities put strong constraints on the energetics and lifetime of the source. Few alternatives without such constraints have been proposed  (e.g. delayed magnetic dissipation proposed by \citet{2006A&A...455L...5G} or photospheric emission from slow material ejected together with the GRB relativistic ejecta \citep{2016MNRAS.457L.108B}). 

On the other hand, the emission of the reverse shock propagating within the ejecta is very sensitive to the structure of the ejecta and may provide an alternative explanation. \citet{2000ApJ...535L..33S}
showed that this reverse shock emission can be long-lasting, if the ejecta ends with a tail of low Lorentz factor material. \citet{2007ApJ...665L..93U,2007MNRAS.381..732G} proposed that such a long-lived reverse shock (LLRS) may dominate the observed emission and could easily produce plateaus. The capacity of this model to reproduce the diversity of the observed X-ray and optical light curves, including plateaus and chromatic breaks was successfully tested by \citet{2011A&A...534A.104H,2012A&A...541A..88H,2012ApJ...761..147U}. Several observed correlations between  the prompt and plateau properties can also be accounted for \citep{2014MNRAS.442...20H}.

\citet{2015arXiv150308333H} suggested that this scenario could also naturally explain flares, alleviating the constraints on the lifetime and variability of the central engine. The flares are associated with the propagation of the LLRS through dense shells within the ejecta, which are expected to appear after the development of internal shocks.  \citet{2015arXiv150308333H}  presented simplified simulations based on a ballistic model \citep{1998MNRAS.296..275D} and computed bolometric light curves assuming an anisotropic emission in the comoving frame \citep{2011MNRAS.410.2422B}. They obtained asymmetric flares with $\Delta t_\mathrm{obs}/t_\mathrm{obs}\simeq 0.1-0.3$. In the present paper, we validate this scenario by performing  relativistic hydrodynamical simulations to compute  the dynamics of the interaction of the LLRS with a structured ejecta.   Although  the analytic models are promising and allow the exploration of a wide range of parameters, only hydrodynamical simulations can account for the full dynamics of the shock interactions and are a necessary step for the validation of the model.

Fully modeling the propagation of a LLRS is a numerical challenge. On one hand, one needs a high enough resolution to properly model the internal shocks. On the other hand, one needs to follow the ejecta beyond the deceleration radius to follow the impact of the LLRS. As such, multidimensional simulations have specifically focused on the internal shock phase \citep[see e.g.][]{2000ApJ...531L.119A,2015ApJ...806..205D}  or the deceleration phase \citep{2007MNRAS.376.1189M,2009A&A...494..879M,2010ApJ...722..235V,2012ApJ...746..122D}. Currently, parameter exploration including both phases can only be achieved with 1D simulations.  In this paper, we perform 1D relativistic hydrodynamic simulations to assert the validity and limits of the ballistic model described in \citet{2015arXiv150308333H}. In \S\ref{sec:setup} we describe our numerical methods and the setup of our set of simulations. In \S\ref{sec:hydro} we analyze the hydrodynamic 
structure of the ejecta and compute the resulting bolometric flux and compare to the ballistic model in \S\ref{sec:emission}. We discuss the observability of X-ray flares in  \S\ref{sec:discussion} and conclude (\S \ref{sec:conclusion}).

\section{Numerical Methods}\label{sec:setup}

\subsection{Relativistic RAMSES}
Our numerical method is based on the relativistic extension of the RAMSES code \citep{2002A&A...385..337T}, presented in \citet{2013A&A...560A..79L}.   The code computes the evolution of the conserved variables in the frame of the laboratory
\begin{equation}\label{eq:cons_prim}
 \mathbf{U}=
 \begin{pmatrix} 
D \\ 
M\\
E 
\end{pmatrix}
=
\begin{pmatrix}
\Gamma \rho \\ 
\Gamma^2 \rho v \frac{h}{c^2}\\ 
\Gamma^2\rho h -P
\end{pmatrix}
\, ,
\end{equation}
where $D$ is the density, $M$ the momentum density and $E$ the energy density. $h$ is the specific enthalpy, $\rho$ is the proper mass density,  $v$ is the fluid velocity, $P$ is the gas pressure. The Lorentz factor is given by $\Gamma=(1-v^2/c^2)^{-1/2}$. Within the relativistic hydrodynamics (RHD) simulation, the equations are solved with all velocities normalized to the speed of light. We neglect the dynamical impact of magnetic fields. While the numerical scheme in our simulation and in \citet{2013A&A...560A..79L} are strictly identical, we use a one-dimensional spherical grid and do not include adaptive mesh refinement in the simulations presented here.

In one-dimensional spherical coordinates, the RHD equations along the radial axis are given by
\begin{eqnarray}
	\frac{\partial D}{\partial t}+ \frac{\partial D v}{ \partial r}    &=&   -\frac{2}{r}D v \nonumber\\ 
	\frac{\partial M}{\partial t}+\frac{\partial (M v+P)}{\partial r }&=& -\frac{2}{r}M v \label{eq:hydro_sphere}	\\ 
	\frac{\partial E }{\partial t}+\frac{\partial M c^2}{\partial r}	&=&-\frac{2}{r} M c^2. \nonumber
\end{eqnarray}
The right hand side of the equations corresponds to the  so-called source terms related to the spherical coordinates.
A passive scalar $s$ is included in the simulations  using $S=s\rho \Gamma$ in the laboratory frame and $F=\rho s v \Gamma$ to compute its flux. As explained below, this allows to distinguish the GRB ejecta from the external medium, or different regions in the ejecta.  Eqs.~ \ref{eq:hydro_sphere} are closed with the following equation of state

\begin{equation}
P=(\gamma-1)(e-\rho)
\end{equation}
where $e$ is the sum of the internal energy and the rest mass energy of the fluid. We assume an ultrarelativistic fluid with $\gamma=4/3$. We use the HLLC Riemann solver and a Minmod slope limiter. Due to the spherical expansion of the ejecta, the density and pressure in the inner part of the ejecta become very small and hard to handle numerically. Therefore, we floor the density and pressure to $10^{-10}$.

\subsection{Moving grid}

Modeling the interactions of the different shocks over time scales comparable to the deceleration timescale is the main numerical challenge in our simulations. As the width of the ejecta  of the GRB $\Delta$ is small compared to the deceleration radius $R_{\mathrm{dec}}$, a static grid is not well fit. Instead, we have implemented a moving grid that follows the motion of the ejecta. The grid has a fixed physical width $L_{\mathrm{box}}$, which is large enough to cover most of the ejecta at the
observer time $t_{\mathrm{obs}}$ we are interested in. At each timestep, we compute the position of the forward shock, defined by the outermost cell where the density is at least a thousand times higher than in the next cell. When the forward shock has reached $R_{\mathrm{edge}}$, we shift the grid towards higher radii. The innermost $n_{\mathrm{shift}}$ cells are suppressed, the contents of all other cells are shifted towards the left and $n_{\mathrm{shift}}$ new cells are created on the right side of the simulation. These cells are filled with the user-defined external medium. In our simulations we arbitrarily set $n_{\mathrm{shift}}=100$ and $R_{\mathrm{edge}}-L_{\mathrm{box}}=100$.  We checked that the exact value of $R_{\mathrm{edge}}$ and $n_{\mathrm{shift}}$ does not impact the outcome of the simulation. 

\begin{figure}
\centering
\includegraphics[width = .4\textwidth]{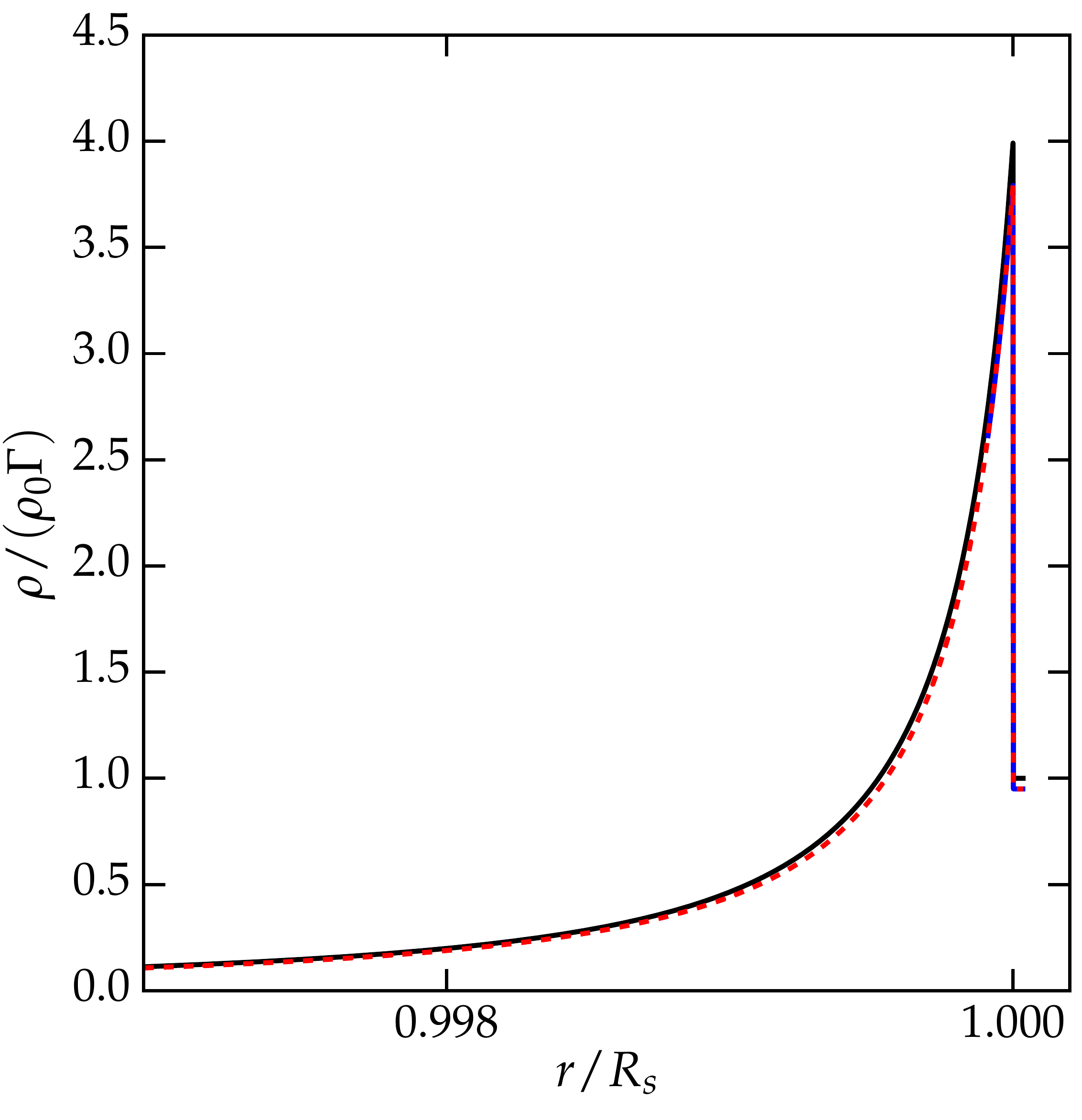}
\caption{\textbf{Validation of the numerical method based on the normalized density in a 1D spherical blast wave.} The simulation on a standard grid (dashed red line), on a moving grid (solid blue line) and the analytic Blandford-McKee solution (solid black line) are in very good agreement, when the forward shock, located at $R_s$ reaches the deceleration radius $R_{\mathrm{dec}}$. The initial energy of the ejecta is $10^{53}$ erg and it propagates in an external medium with $n_0=10^3$\, cm$^{-3}$. At the time of the snapshot, the Lorentz factor behind the shock is 20 and $R_s=4.8\times 10^{16}$ cm.}
\label{fig:BmK} 
\end{figure}

Fig.~\ref{fig:BmK} shows the evolution of a relativistic blast wave as it propagates and expands. The simulation with the moving grid shows the same results as the simulation on the complete grid and is very close to the analytic  Blandford-McKee solution \citep{1976PhFl...19.1130B}. Both simulations have the same equivalent resolution and the moving grid simulation takes roughly ten times less to complete, while modeling most of the mass of the ejecta. While this method perfectly tracks the ejecta over long distances, care has to be taken at the inner boundary.  As such, in this specific simulation we set the boundary condition to match  the analytic solution. For the main simulations presented here, a zero-gradient inner boundary condition is applied. By comparison with a low resolution simulation on a static grid, we find that the zero-gradient inner boundary condition provides a proper result as long as some fraction of the initial ejecta is present in the simulation domain. As such, we stop all our simulations when 50 per cent of the initial ejecta mass has left the domain.

\subsection{Setup of the simulations}

We perform a set of simulations, modeling different levels of complexity in the initial ejecta by varying the initial distribution of the Lorentz factor $\Gamma(r)$ as described below.
The density is then given by 
\begin{equation}
  \label{eq:density}
  \rho(r)=\frac{\dot{E}}{4\pi r^2v(r)\Gamma^2(r)c^2\left(1+\eta\left(\frac{\gamma}{\gamma-1}-\frac{1}{\Gamma(r)^2}\right)\right)}\, ,
\end{equation}
where $\dot{E}$ is the power injected in the ejecta by the central engine and $\eta=10^{-3}$ the ratio between pressure $P$ and the rest mass energy density.  The initial width of the ejecta is set by the duration of the burst $\Delta=ct_{\mathrm{w}}$. In all our simulations we have $t_{\mathrm{w}}=100$\,s, $\dot{E}=10^{51}$ \,erg\, s$^{-1}$.

  The initial inner ($R_{\Delta}$) and outer radii ($R_0$) of the ejecta  are set to $R_{\Delta}/c = $ 100\,s ($\simeq 3 \times 10^{12}$\,cm) and $R_0/c=$ 200\,s\, ($\simeq 6 \times 10^{12}$\,cm). As such, our simulation starts at $t=200$\,s after the start of the relativistic ejection. In all the following, the origin of time is the actual ejection, meaning that our first snapshot is slightly after $200$\,s. As such, we do not model the initial acceleration of the ejecta.

\begin{figure}
\centering
\includegraphics[width = .45\textwidth]{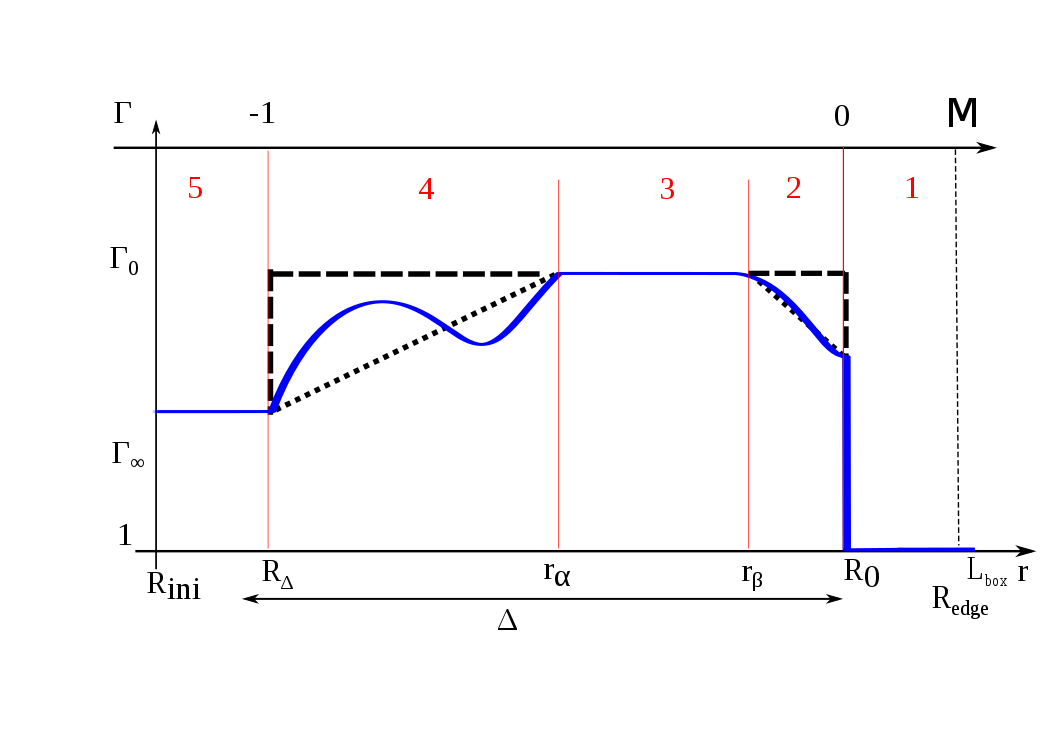}
\caption{\textbf{Initial distribution of the Lorentz factor in the simulations.}  The ejecta extends from $R_{\Delta}$ to $R_0$ and the computational box from $R_{\mathrm{ini}}$ to $L_{\mathrm{box}}$. The distances are not to scale. In the simplest case, the ejecta is uniform (dashed line), while in more complex simulations it includes a tail and/or head region. These outer regions can present a linear variation of the Lorentz factor (dotted line) or present a sinusoidal variation (blue solid line) which will lead to internal shocks.  The vertical red lines show the initial separation of each region in the simulations, with the corresponding values of the passive scalar given in red. The Lagrangian coordinates are indicated on the upper axis.  $R_{\mathrm{edge}}$ is represented by the dashed vertical black line}.
\label{fig:setup} 
\end{figure}

Fig. \ref{fig:setup} provides a schematic view of different initial setups of the Lorentz factor. In the most general case the ejecta displays variability in both the head and tail region (blue solid line) and we have

\begin{eqnarray}
  \label{eq:Lor_ini}
  \Gamma_5(r)&=& \Gamma_{\infty} \nonumber \\
  \Gamma_4(r)&=&\left(\Gamma_0-(\Gamma_{\infty}-\Gamma_0)\frac{r-R_{\Delta}}{\alpha}\right)\left[1+A \sin\left(2\pi \frac{r-R_{\Delta}}{\alpha}\right)\right ] \nonumber \\ 
  \Gamma_3(r) &=&  \Gamma_0 \\
 \Gamma_2(r)&=&  \Gamma_0\left[\frac{1+k}{2}+\frac{1-k}{2}\cos\left(\pi \frac{r-r_{\beta}}{R_0-r_{\beta}}\right)\right] \nonumber \\
\Gamma_1(r) &=&1. \nonumber  
\end{eqnarray}

with 
\begin{eqnarray}
  \label{eq:r_alphabeta}
  r_{\alpha}&=&R_0+(\alpha-1)\Delta \\
  r_{\beta}&=&R_0+(\beta-1)\Delta.
\end{eqnarray}
Table \ref{tab:sims} lists the values of $A,k,\alpha$ and $\beta$ in our simulations. We always adopt $\Gamma_0=100$ and $\Gamma_{\infty}=10$. If the central engine switches off smoothly, one could expect $\Gamma_\infty\simeq 1$. However, this would give a very long deceleration time and radius which are very hard to model numerically. As we are mostly interested in the X-ray variability in the early afterglow phase, our high value is a reasonable approximation. The passive scalar tracks the evolution of different initial regions of the ejecta. The values are indicated on the top row of Fig.~\ref{fig:setup}. The external density is set to $n_0=1000$ cm~$^{-3}$ except for \textit{run5b}, which is identical to \textit{run5} but has $n_0=32$ cm$^{-3}$. The high external density for most of the runs yields a small deceleration radius $R_{\mathrm{dec}}$, making it possible to run our simulations.  As we find no quantitative difference between the outcome of \textit{run5} and \textit{run5b}, we expect the main conclusions of our work to be valid for a more realistic external density.

\begin{table}
\caption{\textbf{Parameters for the simulations.} }
\label{tab:sims}
\centering
\begin{tabular}{c c c c c c c }
\hline
\noalign{\smallskip}
 name &  $A$& $k$ & $\alpha$ &$\beta$ &$n_0$ (cm$^{-3}$) &description  \\ 
\noalign{\smallskip}
\hline
\noalign{\smallskip}
run1  & 0 & 1  & 0   & 1 &$10^3$&uniform                     \\
run2  & 0 & 1  & 0.5 & 1  & $10^3$&tail                  \\
run3  & 0 &0.25& 0.5 & 0.2 &$10^3$&tail+head                   \\
run4  &0.6&1   & 0.5 & 1  &$10^3$&variable tail           \\
run5  &0.6&0.25& 0.5 & 0.2  & $10^3$&variable tail+head          \\
run5b  &0.6&0.25& 0.5 & 0.2  & $32$&variable tail+head          \\
\noalign{\smallskip}
\hline
\end{tabular}

\end{table}

The initial outer edge of the box is located at $L_{\mathrm{box}}/c=600$\,s and we use $N=96000$ grid cells. This gives a final resolution of $\Delta r=1.7 \times 10^8$\,cm, equivalent to a temporal resolution of about 5 milliseconds. Tests with $N=48000$ and $N=192000$ grid cells for \textit{run4} show that the structure of the forward shock is identical in all cases but the internal shocked regions are denser and narrower at higher resolution. However, the Lorentz factor is the same in the whole box for all tests. When comparing the final light curves, we find no difference for the luminosity of the forward shock beyond $t_{\mathrm{obs}}=10$\,s. We find that the total luminosity of the other shocks is higher at higher resolution, with a time-independent offset of about 13 per cent between the $N=48000$ and $N=96000$ runs and a 5 per cent offset between the $N=96000$ and $N=192000$ runs. In all cases, the slopes and variations in the light curves are very similar, indicating that our simulations have enough resolution to study the impact of a LLRS on the light curves.   The parameters we choose are a compromise between the extreme nature of the ejecta and numerical constraints. 

We stop the simulation at the end of the early afterglow phase, about $2\times 10^6$\,s after the initial ejection. At that stage, most of the internal structure of the ejecta has been washed out by shocks and the later evolution would tend to the Blandford-McKee solution \citep{1976PhFl...19.1130B} and then a Sedov solution in the non-relativistic limit \citep{Sedov}.

\subsection{Shock detection algorithm}\label{sec:shock_detection}
To derive light curves from the hydrodynamic structure of the ejecta, we need to determine the energy dissipated in all shocked regions. Common methods to detect shocks include finding important jumps in density and/or pressure or searching for compressed regions ($\nabla \cdot v<0$) \citep{2010MNRAS.404..947C}. Such methods are only successful for very strong shocks, and fail to detect internal shocks in the ejecta.
Therefore, we use a relativistic shock finding algorithm, inspired by \citet{2010A&A...523A...8Z}. The algorithm is based on the predictions of the  wave patterns of the relativistic Riemann problem \citep{2003JFM...479..199R,2002PhRvL..89k4501R}. A shock occurs when the velocity gradient $\Delta v$ between two adjacent cells is greater than a certain threshold $\tilde{v}$, determined by the thermodynamics of the fluid. In practice, for both directions of propagation, we compute the local minima of $V=\tilde{v}-\Delta v$ to determine the locations of the shocks. To avoid spurious detections of very weak shocks, we smooth $V$ over 10 computational cells. This step was not present in \citet{2010A&A...523A...8Z}, who instead suggested to include a small correction to $V$ to avoid spurious detections. 

The determination of the up- and downstream properties of the flow is less precise, as the shocks are smeared out due to numerical diffusion and may interact with each other and the contact discontinuity between the ejecta and the shocked external medium. The spreading of the shocks is most notable at very early times, when shocks are still forming. For a forward/backward propagating shock, we find the cell of maximal density (corrected for the spherical geometry) in the region between 150 and 100 cells before/after the shock location. The downstream density, pressure and Lorentz factor are set to the values in that cell. For the upstream variables, we consider the mean over the cells between 100 and 150 after/before the shock. Visual inspection of the shock locations and conditions confirms the accuracy of the method for $t> 10^5$ s. At  earlier times visual determination of the up-and-downstream regions is  somewhat arbitrary.  However, the hydrodynamic variables we recover respect the Rankine-Hugoniot jump conditions within 10 per cent after $t=10^4$ s. We tested alternate methods to find the downstream variables at earlier times and found no impact on the light curves beyond $t_{obs}=10$\,s. As such, we are confident that our method yields reliable values for the up- and downstream regions and luminosity at the timescales relevant to our study.

The detection of the contact discontinuity between the ejecta and the external medium is made possible with the passive scalar, which values are shown on Fig.~\ref{fig:setup}. For the runs without head region (\textit{run1, run2, run4}) we set the position of the contact discontinuity $r_{CD}$ to be the cell where the passive scalar is the closest to 2. For the runs with head region (\textit{run3, run5}), we set $r_{CD}$ to the cell where the passive scalar is closest to 1.5.

\subsection{Limits of 1D RHD simulations}

The RHD simulations and ballistic model we present are one-dimensional, which enables us to cover a wide range of parameters and study the evolution of the ejecta from the internal shock phase until the deceleration phase. In the ultrarelativistic ejecta we are considering ($\Gamma \geq 10$ at the end of the simulation), the lateral expansion of the jet can be safely neglected and our 1D simulations provide a good model of the global dynamics. Multidimensional simulations show the development of hydrodynamic instabilities \citep{1999ApJ...523L.125A,2013ApJ...775...87D} at the contact discontinuity between the outside medium and the ejecta. While the latter lead to important mixing, they do not strongly affect the structure of the internal and reverse shocks. \citet{2013ApJ...775...87D} show that reverse shocked is pushed away from the forward shock and its emission is more delayed. We do not expect this to qualitatively impact the results presented here. 

Our models also neglect the dynamical impact of the magnetic fields. While the traditional fireball model relies on thermal pressure to accelerate the ejecta, magnetic fields are also promising candidates for the acceleration and collimation of the ejecta during the early phases \citep[see][for a review of magnetic fields in GRB]{2015SSRv..191..471G}. At later stages, when the magnetization is of order unity or above, the propagation of shocks is suppressed \citep{2005ApJ...628..315Z,2008A&A...478..747G,2009A&A...494..879M,2009ApJ...690L..47M,2010MNRAS.407.2501M}. In such case,  one has to invoke magnetic reconnection to explain the observed emission \citep{2011MNRAS.416.2193N,2011ApJ...726...90Z}. As the scenario studied in this paper relies on a complex dynamical evolution related to several generations of shocks, we clearly assume that the ejecta at large distance of the source has a low magnetization ($\sigma\leqslant 0.1$).

In the set of simulations presented here, our assumptions on the initial structure of the ejecta are particularly important, as they lead to an internal structuration after the propagation of internal shocks, which will eventually produce flares in the light curve during the propagation of the LLRS. While we have considered highly idealized cases (see Fig.~\ref{fig:setup}), negative gradients in the Lorentz factor are likely to arise naturally in the flow and a tail of slower material and can be expected from a fading central engine and/or the breakout of the ejecta through the stellar envelope in long bursts \citep[see e.g.][]{2015ApJ...806..205D}. More self-consistent modeling of the flow dynamics would require three dimensional simulations, which come at prohibitive cost for the study presented here.

\subsection{Ballistic model}\label{sec:run1}
If the magnetization is small, the GRB ejecta at large distance from the source is dominated by its kinetic energy. It is then possible to model it using a ballistic approach where pressure waves are neglected. This method is described by \citet{1998MNRAS.296..275D} for the internal shock phase: the outflow is discretized in a large number of shells which interact only by direct collisions. A sequence of collisions models the propagation of a shock wave.
The comparison with a fully relativistic hydrodynamic simulation shows that the ballistic model recovers most features of the evolution \citep{2000A&A...358.1157D}, except for the density in the shocked regions, which is usually underestimated.

This approach has been extended by \citet{2007MNRAS.381..732G,2011A&A...534A.104H,2012A&A...541A..88H,2014MNRAS.442...20H} to include the deceleration phase. This allows to follow the forward shock in the external medium and the reverse shock in the ejecta. The limitations are the same: the radius and Lorentz factor of the different shocks are correctly estimated, as well as the dissipated power, but the density is underestimated. The advantage of the ballistic approach is its very low computational cost, which allows to explore a large domain of parameters for the model. The scenario where afterglow flares are produced by the interaction of a LLRS with dense shells in the ejecta has been explored using the ballistic approach by \citet{2015arXiv150308333H}. The promising results motivated the present study. In the following, we compare our results obtained with the fully relativistic hydrodynamic code described above with the same ejecta modeled with the ballistic approach and identify the different features that will contribute to the light curves.

\section{Hydrodynamics of the ejecta}\label{sec:hydro}

\subsection{Uniform ejecta}
\begin{figure}
\centering
\includegraphics[width = .45\textwidth]{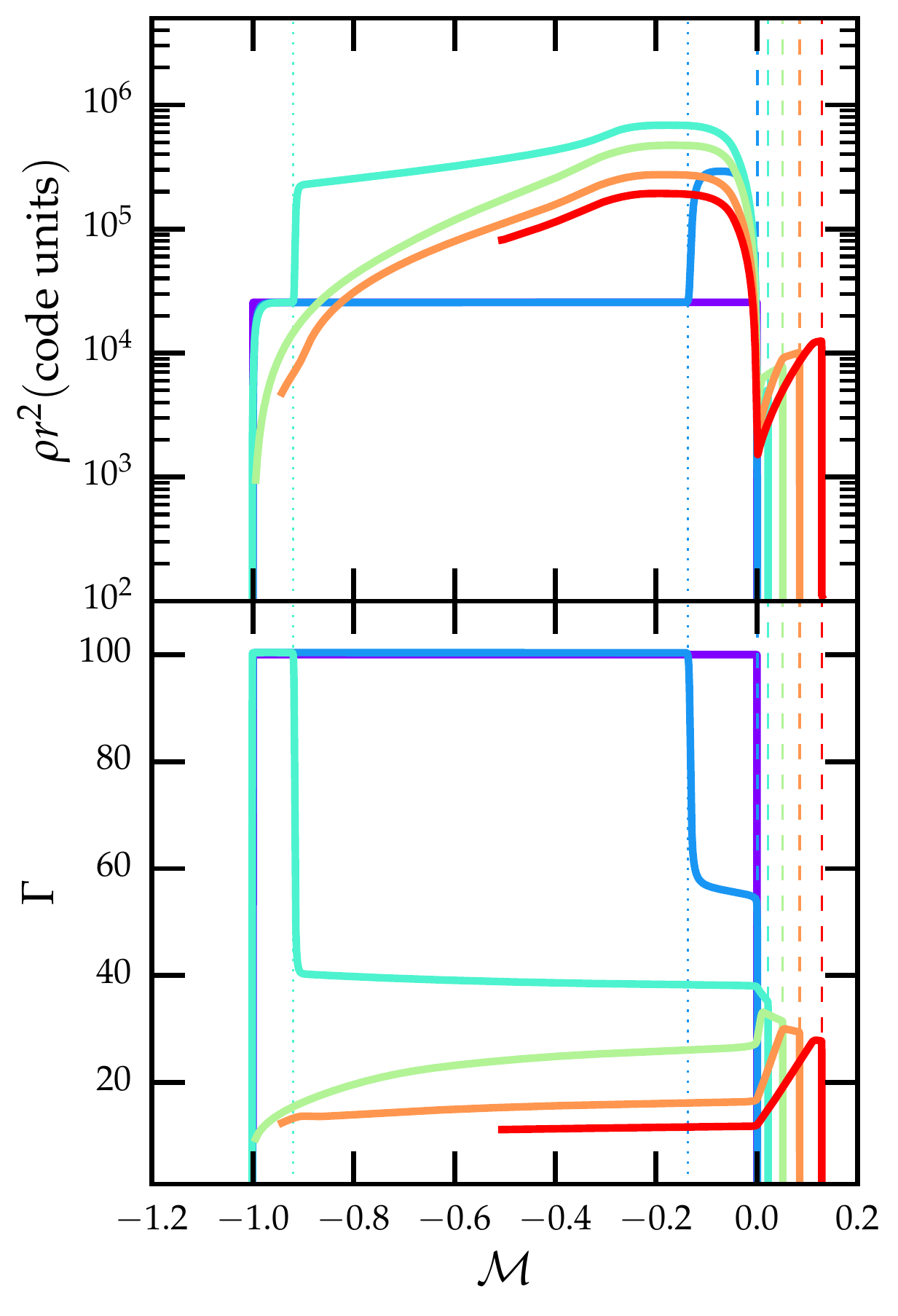}
\caption{\textbf{Hydrodynamics in \textit{run1} with a uniform ejecta}. Density (top) and Lorentz factor (bottom) as a function of the Lagrangian mass at $t=0,2\times 10^5,5.2\times 10^5,6.8\times 10^5, 8.0 \times 10^5 $ and $9.17\times 10^5$\,s (from purple to red). The contact discontinuity between the ejecta and the external medium is located at $\mathcal{M}=0$ by definition. The forward and reverse shock are shown by dashed and dotted vertical lines respectively. We have  multiplied the density by  $r^2$  to remove the impact of spherical expansion and allow a better focus on shocked and expanding regions. After $t=6.8\times 10^5$ s the inner part of ejecta is out of the simulation domain.}
\label{fig:run1} 
\end{figure}

Fig. \ref{fig:run1} shows the density and Lorentz factor in \textit{run1},which we consider as our reference model. The initial state (purple line) corresponds to a uniform shell of width $ct_\mathrm{w}=3.0\times 10^{12}$\, cm with a Lorentz factor $\Gamma_0=100$, an energy $E_0=10^{53}$\,erg and a mass $M_0=\frac{E_0}{\Gamma_0 c^2}=1.1\, 10^{32}$\, g. After describing the evolution in this simple case, we will progressively describe and explain the impact of various features in the setup on the dynamical evolution of the ejecta. We use a Lagrangian description to facilitate identification of different discontinuities. The x-axis shows  the accumulated mass fraction normalized by the initial mass of the ejecta:
\begin{equation}
\mathcal{M}(R)=\frac{1}{M_0}\int_{r_{CD}}^R 4\pi r^2\, \rho \Gamma dr\, .
\end{equation}
As such, we use negative values for the ejecta and positive values for the shocked external medium.

With our initial conditions, the spreading radius of the ejecta is of the order of $R_\mathrm{spread}\simeq\Gamma_0^2 \Delta \simeq 3.0\times 10^{14}$\, cm and the deceleration radius $R_\mathrm{dec}\simeq\left(\frac{3}{4\pi}\frac{\dot{E}t_\mathrm{w}}{\Gamma_0^2 n_0 m_\mathrm{p}c^2}\right)^{1/3}\simeq 1.2\times 10^{16}$\,cm. The condition $R_\mathrm{spread}\ll R_\mathrm{dec}$ corresponds to the case where the reverse shock is initially non-relativistic, becomes progressively relativistic, and crosses the ejecta at $R_\mathrm{cross}\simeq R_\mathrm{spread}^{1/4}R_\mathrm{dec}^{3/4}\simeq 4.8\, 10^{15} $\, cm $\simeq 0.4 R_\mathrm{dec}$, before the deceleration radius \citep{1995ApJ...455L.143S,1999ApJ...513..669K}. This is indeed observed in our simulation. After $t=2\times 10^5$\,s, i.e. $R\simeq 0.5\,R_\mathrm{dec}$ (blue line), the forward shock is still very close to the contact discontinuity and cannot be distinguished on the plot. On the other hand, the reverse shock has already crossed almost 20 per cent of the mass of the ejecta. 
At $t=5.2 \times 10^5$\,s, i.e. $R\simeq 1.3\, R_\mathrm{dec}$, the reverse shock has almost finished to cross the ejecta, and the forward shock has just started to propagate: the shocked external medium has a mass of a few per cent of the initial mass of the  ejecta, i.e. is of the order of $M_0/\Gamma_0$ (dark green line).

After $t\simeq 8\times 10^5$\,s, i.e. $R\simeq 2\, R_\mathrm{dec}$, the forward shock has accumulated a significant amount of external medium, and conversely, some of the initial ejecta has now left the simulation box on the left side (orange line). At $t=9.17\times 10^5$\,s, 50 per cent of the initial  mass of the ejecta has left the simulation domain (red line). The subsequent evolution is only driven by the forward shock and we recover the temporal evolution from analytic solution of \citet{1976PhFl...19.1130B}. However, as the shocked external medium accumulates more mass and becomes wider, it cannot be modeled properly within the size of our simulation box. After $t\simeq 1.8\times 10^6$\,s, i.e. $R\simeq 4.5\,R_\mathrm{dec}$,
our simulation diverges from the \citet{1976PhFl...19.1130B} and we discard its subsequent evolution. 

\subsection{Long-lived reverse shock}

\begin{figure}
\centering
\includegraphics[width=.45\textwidth]{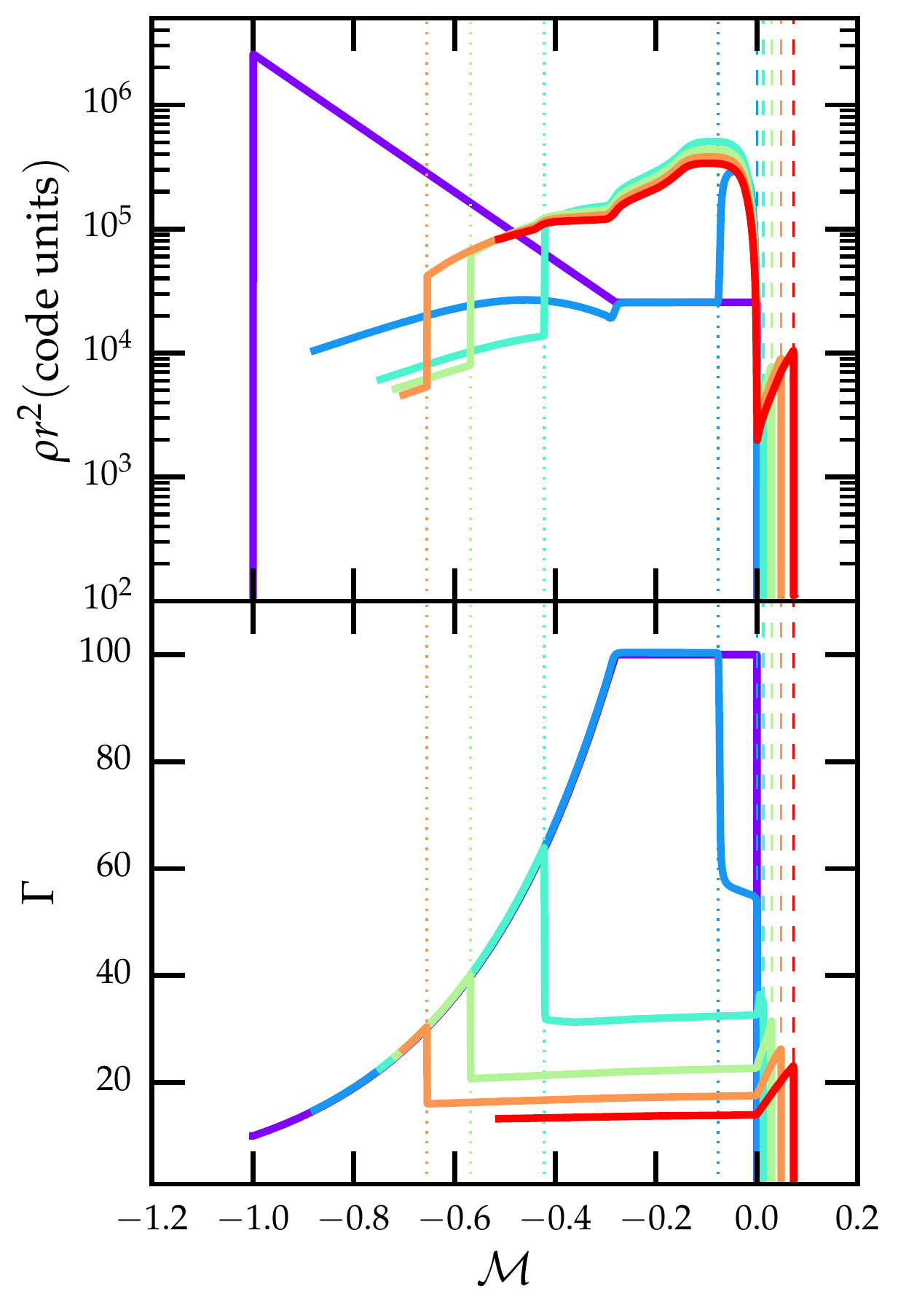}
\caption{\textbf{Hydrodynamics in \textit{run2}, which  includes a slower tail region.} The timesteps and colors are the same as in Fig.~\ref{fig:run1}. }
\label{fig:run2} 
\end{figure}

The luminosity of the afterglow  may be altered if the reverse shock is long-lived. Therefore, we reduce the speed of the second half of the ejecta, with a linearly decreasing Lorentz factor in \textit{run2}. Fig.~\ref{fig:run2} shows the evolution of \textit{run2}, at the same times as \textit{run1}. As the energy injection in both ejecta is set to be identical, the total mass in the ejecta in \textit{run2} is about twice as large than in \textit{run1}. The gradient in the Lorentz factor causes the ejecta to expand and form a tail of slow material, and a fraction of the ejecta is rapidly advected out of the simulation domain. As expected for this initial state, we observe a LLRS: (i) before the reverse shock reaches the tail of the ejecta around $t=4\times 10^5$\,s
(transition happens between blue and green lines), the evolution downstream of the reverse shock is exactly the same as for \textit{run1}; (ii) afterwards, both runs evolve differently. Due to the slower and denser tail region, the propagation of the reverse shock is slowed down with respect to the uniform case and it leaves the simulation region  at $t=8.7 \times 10^5$\,s (compared to $5.5\times 10^5$\,s in \textit{run1}, transition happens between orange and red lines). Similarly, the slower tail region results in a slower forward shock.

\subsection{Internal shocks}
Internal shocks occur when the gradient of the Lorentz factor is strictly negative meaning faster material is behind slower material. We explore two different cases: when a head region is present ahead of the fastest part of the ejecta as in \textit{run3} or when a sinusoidal variation in the Lorentz factor is present in the tail region as in \textit{run4}.  In both cases, we expect internal shocks to form and structure the ejecta, but the interaction of these shocks with the reverse shock should occur much earlier in \textit{run3}. This is confirmed by the simulations.

\begin{figure*}
\centering
\includegraphics[width=.454\textwidth]{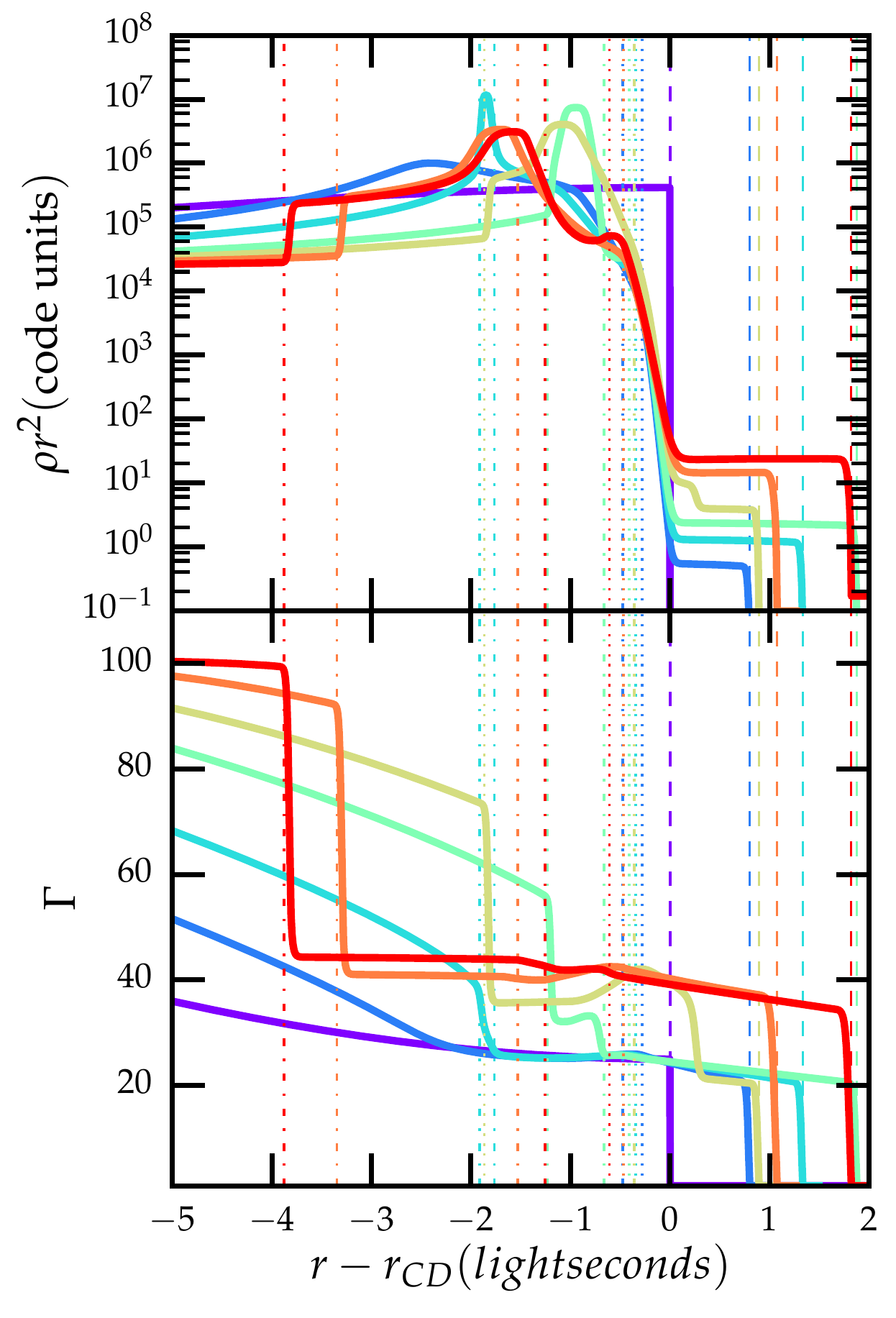}
\includegraphics[width=.45\textwidth]{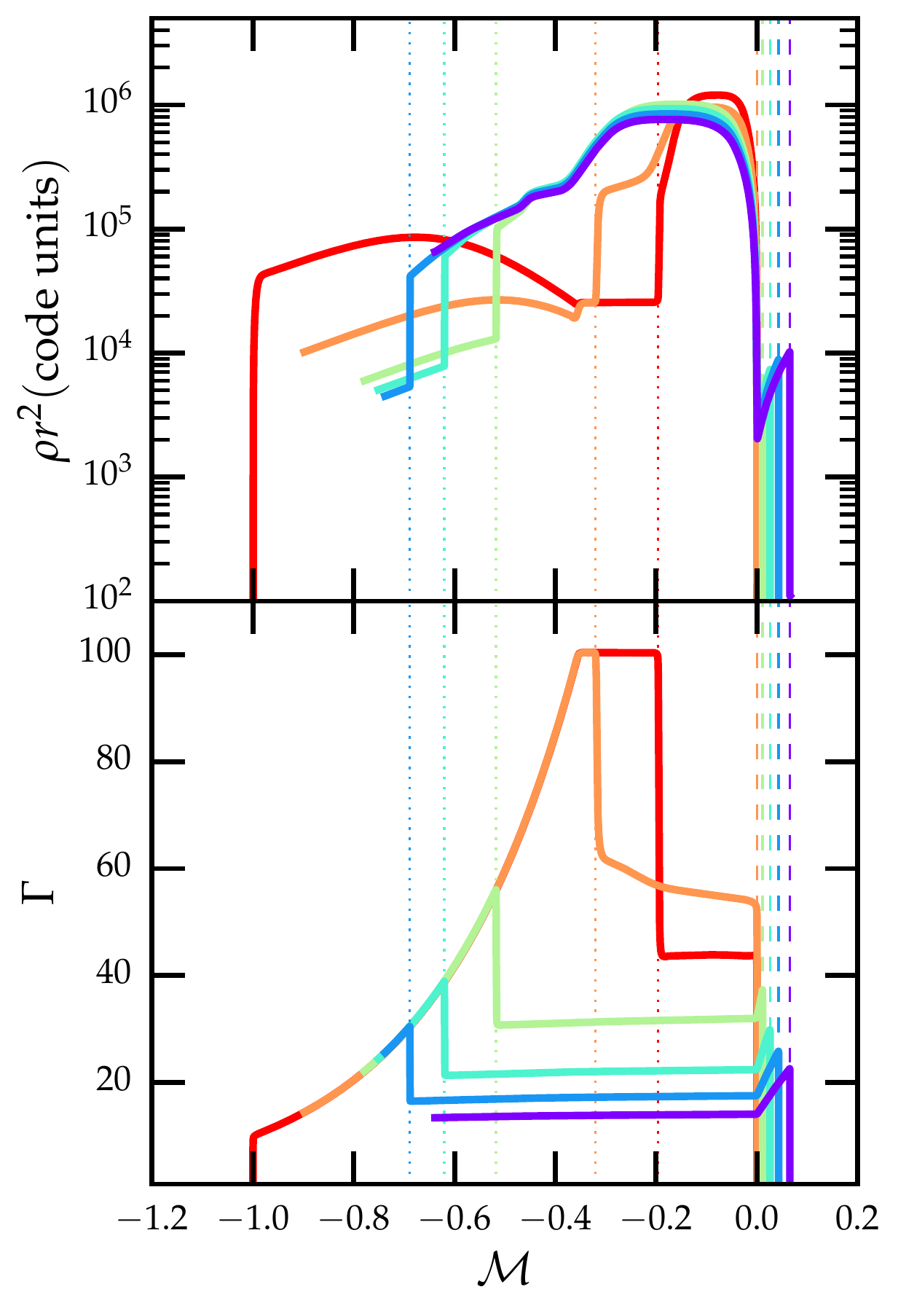}
\caption{\textbf{Hydrodynamics in \textit{run3}, which includes a slower head region.} Left: Eulerian view of the early evolution of the density and Lorentz factor at $t=0,6\times 10^3,1\times 10^4,1.4\times 10^4,1.8\times 10^4,2.7\times 10^4$ and $3.6\times 10^4$\,s (from purple to red). The dotted, dot-dashed and dashed lines respectively show the position of the reverse, internal and forward shocks. Right: Lagrangian view of the later evolution, showing the same quantities at $t=3.6\times 10^4$\,s, which is the latest timestep on the left panel (red), and $2\times 10^5$, $5.2\times 10^{5}$, $6.8\times 10^{5}$, $8.0\times 10^{5}$, $9.17\times 10^{5}$ (from orange to purple). The timesteps on the right are the same as in Figs.~\ref{fig:run1},\ref{fig:run2}.}
\label{fig:run3}
\end{figure*}

Fig.~\ref{fig:run3} shows the evolution of \textit{run3}, with the left panel focusing on the early development of an internal shock. We used Eulerian coordinates as they better highlight different discontinuities in the head region at this early stage. Initially, the Lorentz factor profile is very shallow, but it quickly steepens as faster material is behind slower material. We expect the formation of internal shocks at a radius $R_\mathrm{is}\simeq 2\left(k \Gamma_0\right)^2 c t_\mathrm{var}$, where $t_\mathrm{var}$ is the variability timescale in the initial state, of the order of $0.5\, \beta t_\mathrm{w}$, i.e. $R_\mathrm{is}\simeq 4\, 10^{14}\, \mathrm{cm}$. Indeed, we observe  that after roughly $10^4$\,s, a denser shocked region has developed and two internal shocks appear (dot-dashed lines). However these shocks do not have time to propagate as, at $t=1.5\times 10^4$\,s, 
the reverse shock reaches the outer edge of the dense shocked region (transition happens between green and yellow lines).
This results in a reflected shock propagating forward, as can be seen at $t=1.8\times 10^4$\,s (yellow line). This internal shock crosses the contact discontinuity at $t=$ 1.6$\times 10^4$ s (transition happens between yellow and orange lines) and eventually catches up with the forward shock in the external medium at $t=2\times 10^4$\,s, reaccelerating it. The later propagation of the reverse shock is very similar to \textit{run2}.

\begin{figure*}
\centering
\includegraphics[width=.45\textwidth]{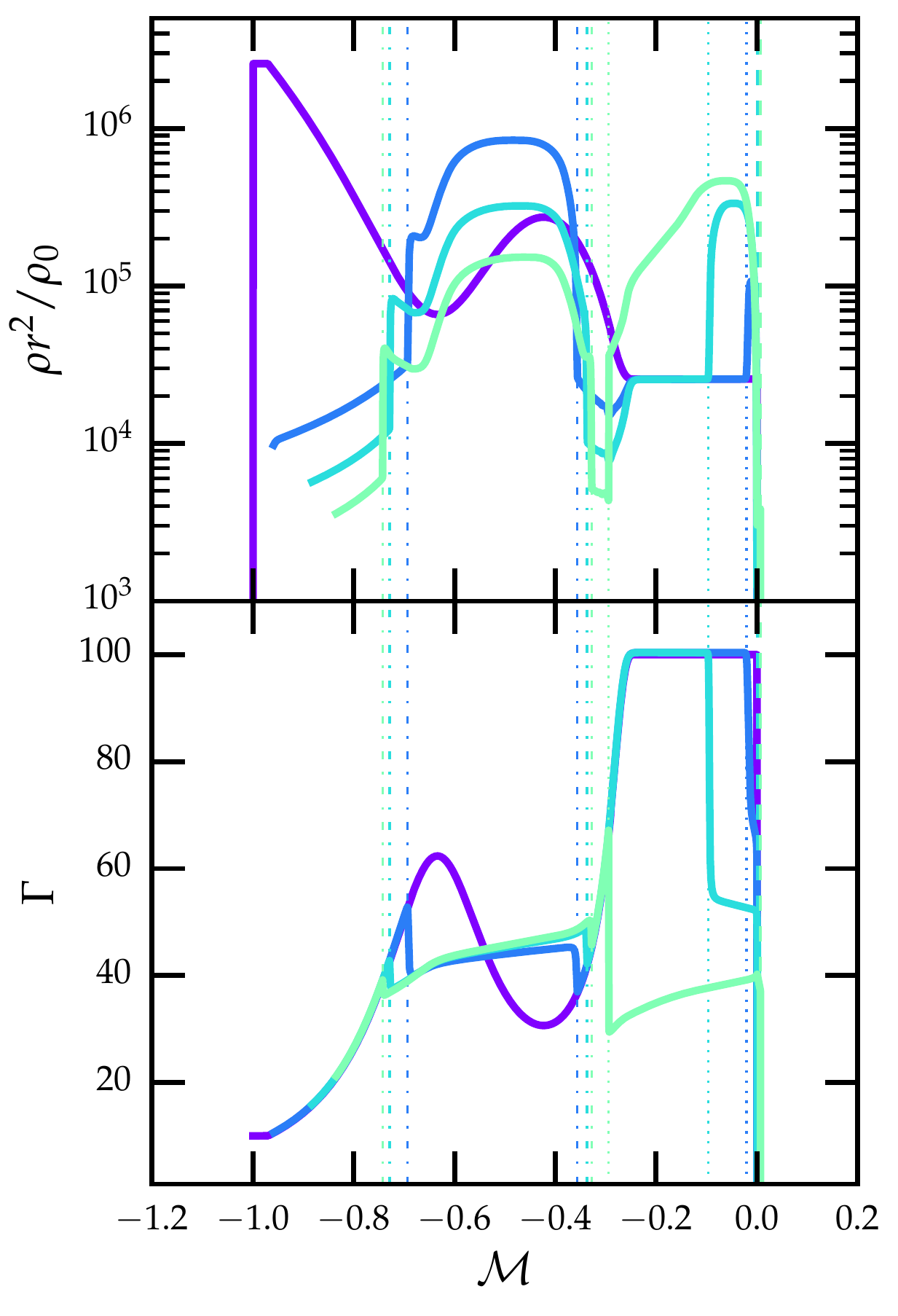}
\includegraphics[width=.45\textwidth]{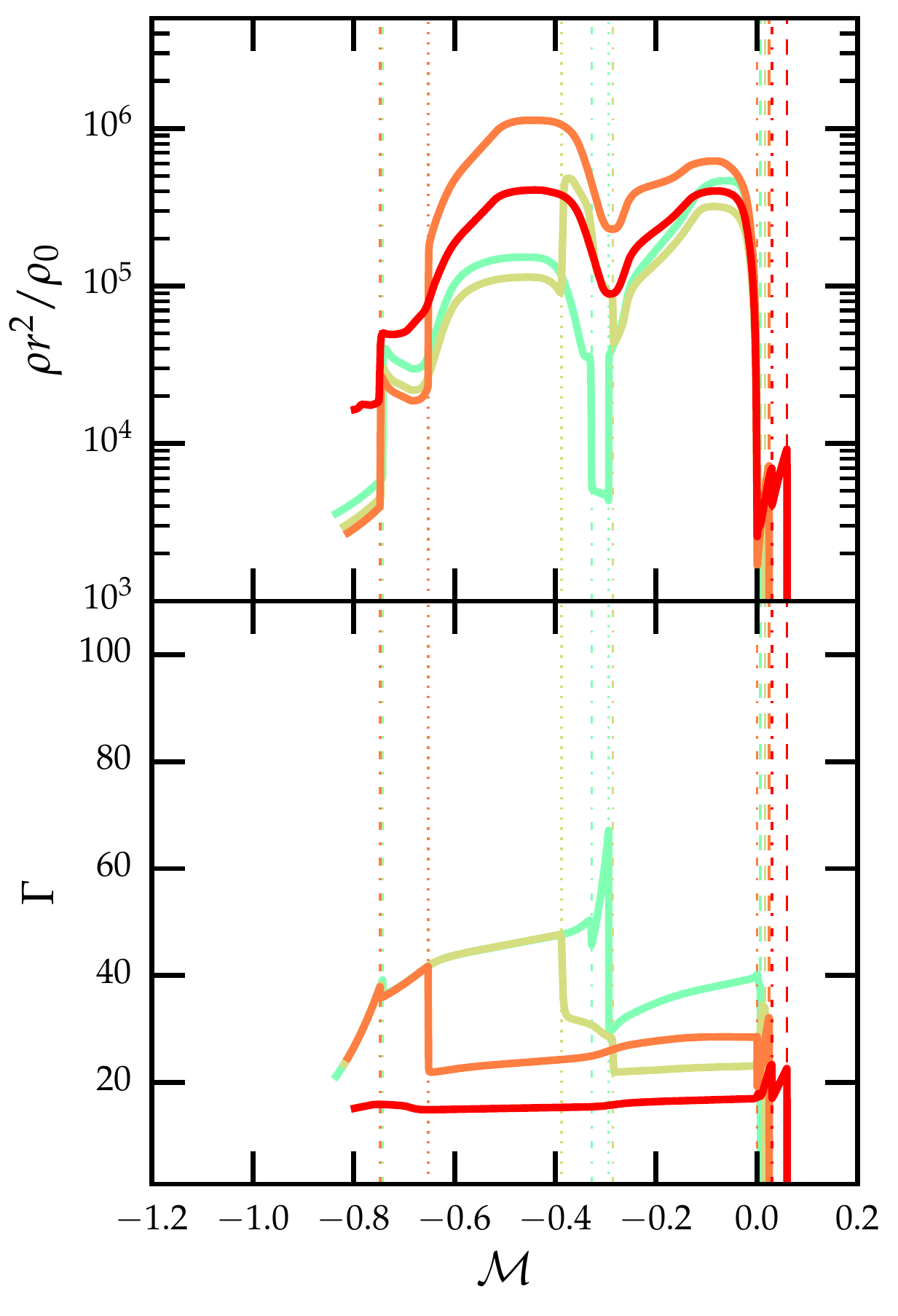}
\caption{\textbf{Hydrodynamics in \textit{run4}, which  includes variability in the tail region. }Left: Early evolution showing $t=0,10^5,2.3\times 10^5$ and $4.4\times 10^5$\,s (from purple to green). Right : Later evolution showing $t=4.4\times 10^5, 5.7\times 10^5, 6.5\times 10^5$ and $8.7 \times 10^5 $\,s (from green to red). }
\label{fig:run4}
\end{figure*}

In \textit{run4}, the negative Lorentz factor gradient is located in the tail region, where is has more time to develop into a dense shocked region before it becomes affected by the reverse shock at $t=5\times 10^5$\,s. The left panel of Fig. \ref{fig:run4} shows this initial development, until $t=4.4\times 10^5$\,s. As the reverse shock interacts with the outermost internal shock, a shock reflection occurs and results in a forward propagating shock (see yellow line), similarly to \textit{run3}. The latter will propagate into the shocked external medium (see red line) and eventually catch up with the forward shock. The propagation of the reverse shock lasts for almost $10^6$\,s, as it is significantly slowed down when crossing the dense region. When it eventually crosses the backward propagating internal shock, a reflected shock propagates forward again, which also eventually reaches the forward shock and slightly re-accelerates it (see red line). This evolution is in good agreement with the theoretical expectation described in \citet{2000ApJ...532..286K} and section 4 of \citet{2015arXiv150308333H}.
As described in \citet{2015arXiv150308333H}, this interaction of the reverse shock with the dense shell in the tail due to the formation of internal shocks will affect the luminosity and is a possible mechanism for observed flares in the afterglow.

\subsection{Full evolution}
Having identified the individual contributions of each feature, we analyze \textit{run5}, which combines all previous features and provides a more complete insight on the evolution of a realistic GRB afterglow. The global evolution is  essentially a combination between the impact of the early internal shocks present in \textit{run3} and the later ones from \textit{run4}. This setup leads to two internal shock regions, which both have a forward propagating shock and backwards propagating shock. The interaction between the latter and the reverse shock leads to reflected shocks, which propagate back to the forward shock.

\begin{figure*}
\centering
\includegraphics[width=.9\textwidth]{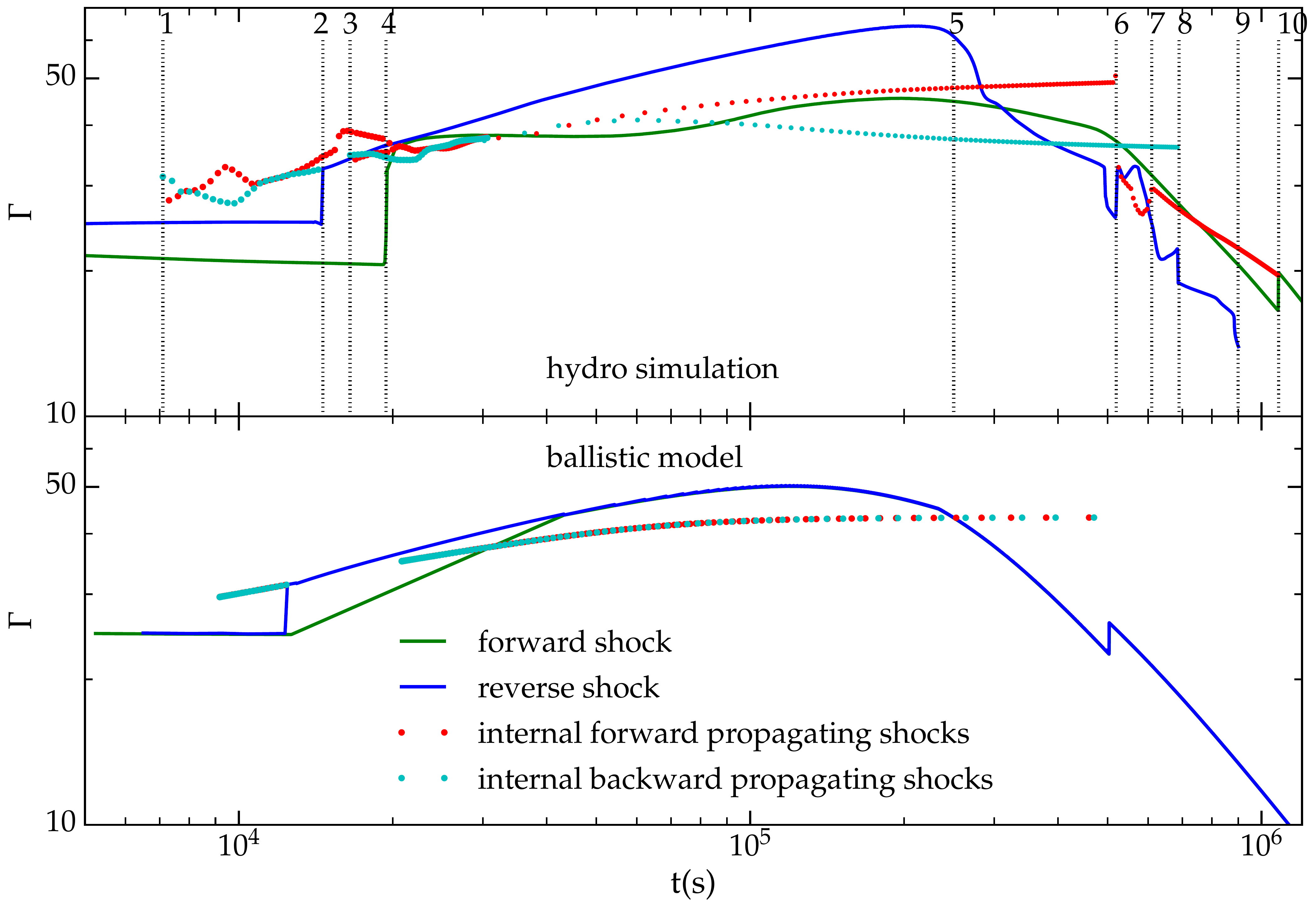}
\caption{\textbf{Lorentz factor in \textit{run5}, which includes variability in the tail region and a slower head region.} The upper and lower panels show the Lorentz factor in the shocked material behind the forward  shock (FS), reverse shock (RS) and the internal forward (IFS) and backwards propagating shocks (IRS) in the hydrodynamic simulation and ballistic model respectively. Sudden variations are indicated with numbers and vertical lines and are detailed in the text. }
\label{fig:run5}
\end{figure*}

Fig.~\ref{fig:run5} shows the Lorentz factor downstream each of these shocks over time, with the results of the ballistic model for the same initial state shown below. At $t=7\times 10^3$\,s, the internal shocks in the head region become apparent (label 1 in Fig.~\ref{fig:run5}) and persist until the reverse shock interacts with its forward propagating shock (2). This effectively damps the reverse shock, and the internal reverse shock is now considered as the reverse shock. A reflected shock is  swiftly propagating forward until it catches up with the forward shock (4) and reaccelerates it significantly.  In the meantime, a shocked region has formed in the tail region as well (3).  As the reverse shock enters the tail region it is slowed down (5) until it is accelerated when encountering the internal shock region (6). Again, this interaction results in a forward propagating shock which is momentarily stalled at the contact discontinuity with the shocked external medium (7) and then catches up with the forward shock (10). The reverse shock eventually reaches the back shock of the internal shocked medium (8) and then leaves the simulation domain (9). In the next section we study the impact of the dynamics on the bolometric light curves of the ejecta.

Fig.~\ref{fig:run5} also highlights that in the ballistic model, the shocked regions behind the reverse shock and the forward shock have the same Lorentz factor, as behind the internal forward and reverse shocks. This is due to the lack of resolution in this approach, where shells merge after each collision. Generally, the Lorentz factors in the simulations bracket the value from the ballistic model. The Lorentz factor of the forward shock is usually underestimated while the reverse shock is modeled more accurately. Conversely, the downstream density in the ballistic model (not shown here) seems well reproduced for the forward shock, but strongly underestimated for the reverse shock. In the next sections we compare the resulting bolometric luminosity of both models.

\section{Bolometric emission}\label{sec:emission}
In this section we compute bolometric lightcurves based on the different runs. The advantage is that this calculation is independent on any assumption on the microphysics in the various shocked regions. It shows where the power is dissipated and  provides an indication of the shape of the observed emission at frequencies dominated by fast-cooling electrons, as it is expected for X-rays at early times. In addition, the bolometric lightcurves allows a direct comparison with the simulations made with the ballistic model. In \S\ref{sec:discussion}, we will introduce microphysics parameters to
provide a more specific discussion of the X-ray emission to explore the validity of a X-ray flare model based on a long-lived reverse shock in a stratified ejecta.
\subsection{Simplified emission model}
We compute the bolometric light curves for our different models and determine the contributions of the forward shock, internal shocks and reverse shock. At a given \tobs in the observer frame, the received bolometric luminosity from one shocked region is given by \citep{1999ApJ...523..187W}
\begin{equation}
  \label{eq:lbol}
    L_{\mathrm{bol}}(t_{\mathrm{obs}})=\int \mathrm{d}t  \frac{2 \epsilon_e \, A(\theta)\, L_{\mathrm{diss}}(t) }{\Delta t_{\mathrm{obs}}(t)} \frac{1}{\left(1+\frac{t_{\mathrm{obs}}-t_{\mathrm{obs},0}(t)}{\Delta t_{\mathrm{obs}}(t)}\right)^3}\, ,
\end{equation}
where the integration over $t$ (lab frame) is carried out during the time interval corresponding to the propagation of the considered shock. In Eq.~(\ref{eq:lbol}), $L_{\mathrm{diss}}(t)$ is the dissipated power at the  shock, $\epsilon_\mathrm{e}$ is the fraction of the dissipated energy injected in non-thermal electrons,
\begin{equation}
  \label{eq:tobs0}
  t_{\mathrm{obs},0}=t-\frac{r(t)}{c}\, 
\end{equation}
is the observer time for the reception of the first photons emitted on-axis at time $t$ and radius $r(t)$, 
and 
\begin{equation}
  \label{eq:dtobs}
\Delta t_{\mathrm{obs}} (t)=\frac{r(t)}{2\Gamma^2 c}   \, 
\end{equation}
is the observed time delay between photons emitted at time $t$ either on-axis or at an angle $1/\Gamma$. 

The dissipated energy at a given shock $L_{\mathrm{diss}}(t)$ is set by the variation of the internal
energy between the downstream and upstream regions 
\begin{eqnarray}
  \label{eq:Ldiss}
  L_{\mathrm{diss}}(t)&=&\dot{M}_s \Gamma_*(\epsilon_*-\epsilon)\\
  &=& 4\pi r^2  \rho_* \Gamma_*^2(V_s-v_*)(\epsilon_*-\epsilon),
\end{eqnarray}
 where the subscript $*$ indicates quantities in a shocked region, $\dot{M}$ is the mass flow across the shock, $\epsilon$ the specific internal energy and $V_s$ is the velocity of the shock, which is determined by comparing its position between outputs. 

Compared to \citet{1999ApJ...523..187W}, the function $A(\theta)$ has been introduced in Eq.~(\ref{eq:lbol}) by \citet{2011MNRAS.410.2422B} to account for a possible anisotropy of the synchrotron radiation in the comoving frame. $\theta$ is the angle with respect to the radial direction in the comoving frame, such that
\begin{equation}
\cos{\theta}=\frac{
\Delta t_{\mathrm{obs}}(t)
-\left(
t_{\mathrm{obs}}-t_{\mathrm{obs},0}(t)
\right)
}{
\Delta t_{\mathrm{obs}}(t)
+\left(
t_{\mathrm{obs}}-t_{\mathrm{obs},0}(t)
\right)}\, .
\end{equation}

In this section, we limit our study to the bolometric light curve in the simplest case, where the emission in the comoving frame is isotropic, i.e. $A(\theta)=1$. In the early X-ray afterglow and during flares, the radiating electrons are expected to be in fast-cooling regime, and therefore the bolometric lightcurve already gives a fair idea of the shape of the predicted emission. We introduce microphysics parameters and discuss the effect of anisotropy in \S\ref{sec:discussion}.
 In practice, for each output (corresponding to a given time $t$), 
 we determine the location of the different shocks,  and then for each of them determine $t_{\mathrm{obs},0}$ and $\Delta t_{\mathrm{obs}}$ and the quantities in the shocked region to compute the corresponding $L_{\mathrm{diss}}$.
 To compute the bolometric light curve due to a given shock, for each $t_{obs}$, we add the contributions of all $t$ using Eq.~\ref{eq:lbol}. We  eventually add up the contributions of the different types of shocks. 

\subsection{Bolometric light curves}

\begin{figure*}
\centering
\includegraphics[width=.8\textwidth]{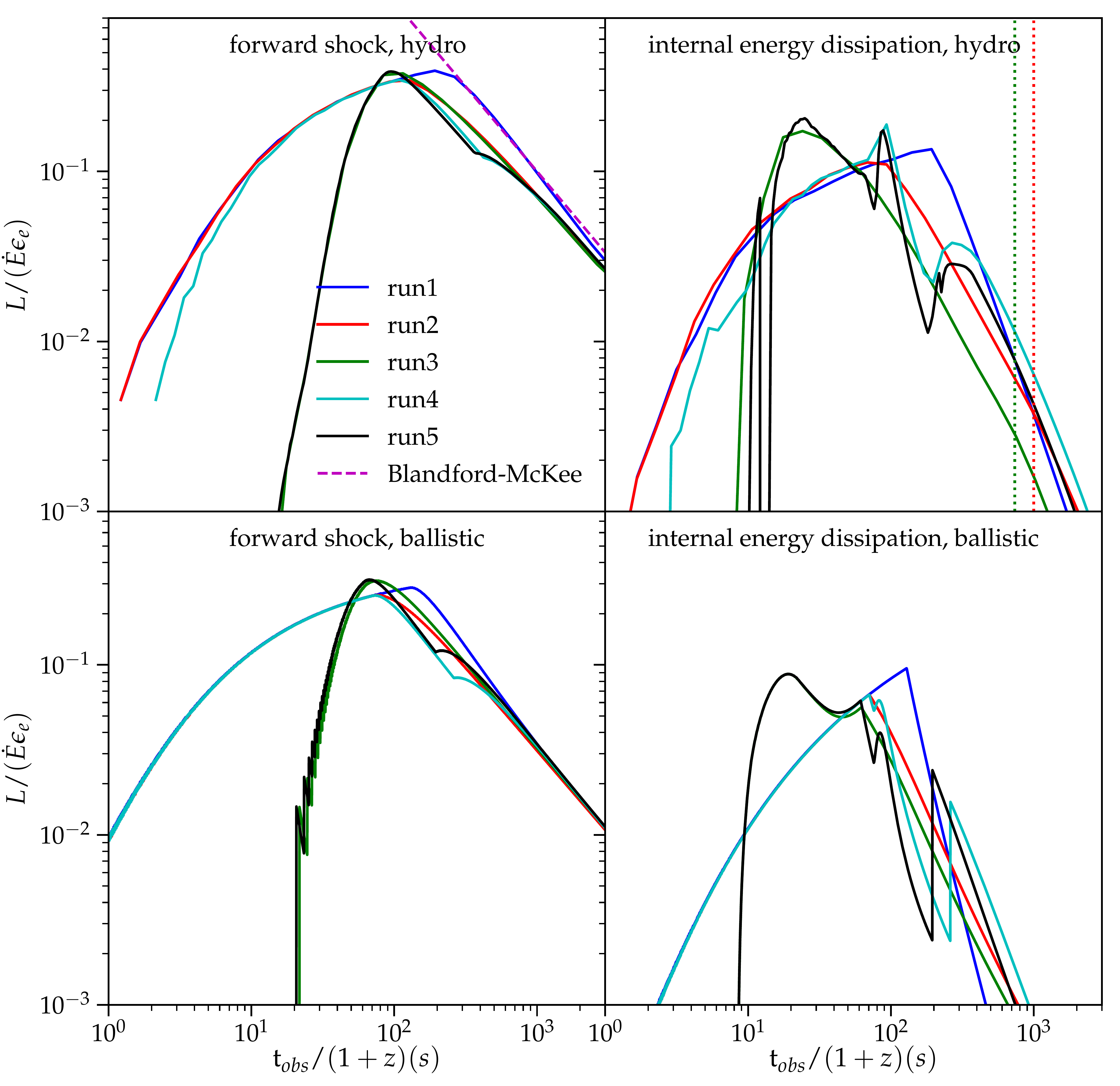}
\caption{\textbf{Bolometric light curves of the forward shock (left column) and  internal and reverse shocks (right column) assuming isotropic emission in the comoving frame.}  The five main runs are shown (color coded), comparing the simulations (upper row) and the ballistic model (lower row). On the top-left panel, the magenta dashed line shows the $t_{\mathrm{obs}}^{-1}$ \citep{1976PhFl...19.1130B} self-similar solution. The dotted vertical  lines indicate the time beyond which some of the off-axis emission is missing because the reverse shock has left the simulation domain. As seen in the right column, internal dissipation  leads to flare-like features for $t_{obs}\geqslant 70$ s  in \textit{run4} and \textit{run5}. }
\label{fig:lightcurves}
\end{figure*}

 Fig.~\ref{fig:lightcurves} shows the bolometric light curves for all runs and the result of the ballistic model. The left column shows only the contribution of the forward shock propagating in the external medium. In \textit{run1}, its 
 bolometric luminosity initially increases until the head of the ejecta starts decelerating at $t_{\mathrm{obs}} \simeq 200$\,s, after what it follows the expected self-similar solution with a $t_{\mathrm{obs}}^{-1}$ slope \citep{1976PhFl...19.1130B}. When the head of the ejecta is slower (\textit{run3, run5}), the initial luminosity is naturally lower and suddenly rises after $t_{obs} =20$~s when the forward shock is reaccelerated by the reflected forward shock (label '4' in Fig.~\ref{fig:run5}).  The forward shock is more luminous than for \textit{run1} for a brief moment and the deceleration occurs earlier. Eventually, all runs converge to the same self-similar solution, consistent with the total amount of injected energy.  The bump in \textit{run4, run5} around $t_{\mathrm{obs}}\simeq 300$\,s is caused by the reacceleration of the forward shock due to the internal forward propagating shock (the latter resulting from the reflection of the reverse shock on the internal shock region, label '10' in Fig.~\ref{fig:run5}). The luminosity curves are in good agreement with the ballistic model shown below. In all cases, the simulations yield a slightly later deceleration of the ejecta.

The right column of Fig.~\ref{fig:lightcurves} shows the sum of the contributions of all other shocks, which propagate within the relativistic ejecta. This includes the reverse shock, and the internal shocks in the tail and head regions, and represents the total internal dissipation. As suggested by the complex evolution of the Lorentz factor in Fig.~\ref{fig:run5}, the bolometric light curves are not smooth and present a much higher variability than the lighturves from the forward shock. This important variability, which will be fully described in \S\ref{sec:discussion}, confirms our initial motivation for this study.

For \textit{run1}, the luminosity of the reverse shock progressively increases until it reaches the inner edge of the ejecta at $t=5.2\times 10^5$\,s, corresponding to $t_{\mathrm{obs}}=200$\,s, after which only off-axis photons are received and the flux follows the expected $t_{\mathrm{obs}}^{-3}$ decrease.
For \textit{run2}, which includes a slower tail region and yields a LLRS, the decrease in luminosity starts earlier, when the reverse shock reaches the tail region. As the ejecta is more extended in this case, the reverse shocks remains active for a longer time and the luminosity ends up being larger than for \textit{run1}. In both cases, the ballistic model very well reproduces the simulation, with a slightly lower and steeper drop-off for the luminosity.

When the head region is present, as in \textit{run3} and \textit{run5}, the reverse shock is slower, and the emission very steeply rises around $t_{\mathrm{obs}}=10-20$\,s, when the reverse shock encounters the shocked region (label '2' in Fig.~\ref{fig:run5}). The following decrease in flux is initially slow but recovers the same slope than \textit{run2} after roughly 100\,s. 
The small upturn seen around $t_{\mathrm{obs}}=70$\,s is not recovered in the ballistic model.

In \textit{run3} and \textit{run5}, the impact of the internal shocks in the head region is rapidly suppressed by the propagation of the reverse shock and  this configuration is unlikely to produce important variability\footnote{However, variations of the Lorentz factor on shorter timescales would lead to earlier internal shocks, which could propagate without being smoothed out by the reverse shock. As our study does not focus on the prompt phase, we did not simulate such cases. } On the other hand, internal shocks in the tail region produce a much stronger effect, as can be seen in \textit{run4}. The initial evolution is identical to \textit{run2} and the impact of the internal shock region becomes apparent around $t_{\mathrm{obs}}=100$\,s,  just after the dimming of the reverse shock as it enters the tail region. As the reverse shock interacts with the internal shocked region, it gets revived and results in the reflected forward propagating shock. Both these effects account for the rebrightening at $t_{\mathrm{obs}}\simeq 250$\ s, with the main contribution from the reflected shock. The ballistic model shows qualitatively similar results, although with lower emission at all times.

\begin{figure*}

\centering
\includegraphics[width=.45\textwidth]{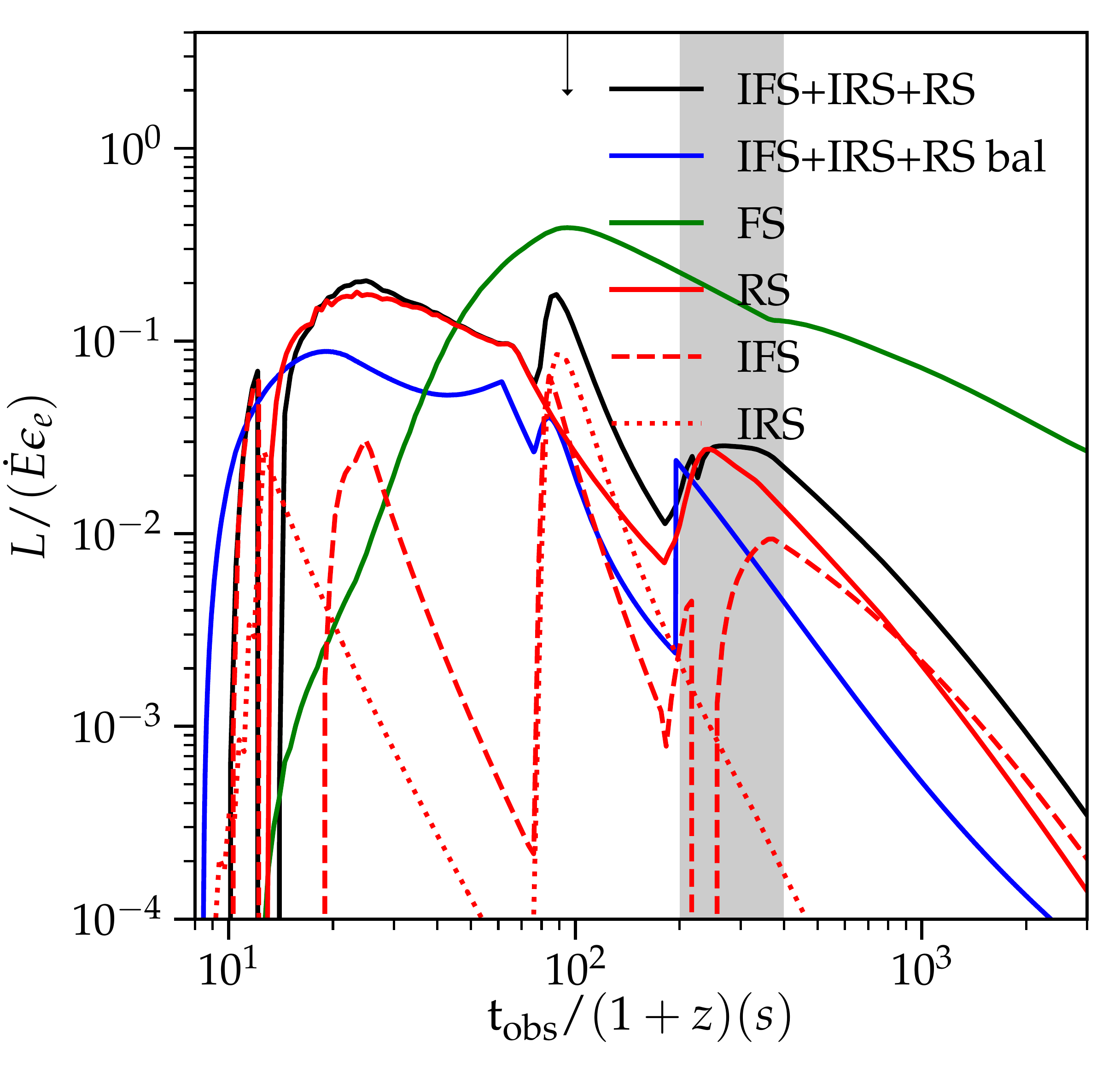}
\includegraphics[width=.45\textwidth]{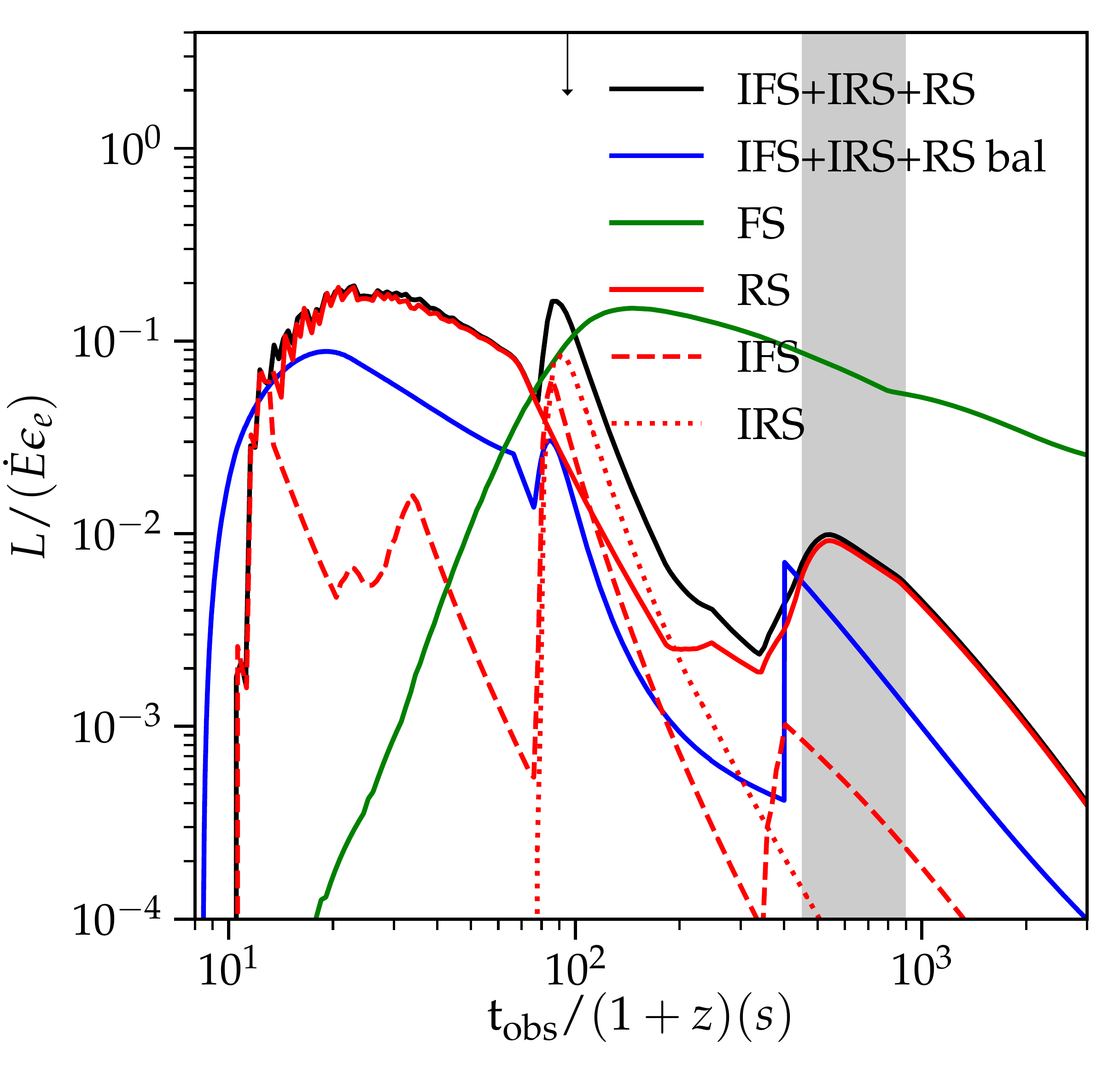}
\caption{\textbf{Contribution to the bolometric lightcurves of the different shocks in \textit{run5} (left) and \textit{run5b} (right).} We separately show the energy dissipation at the forward shock (FS) and the sum of all the internal dissipation (IFS+IRS+FS), and compare with the ballistic model (blue). The flares and rebrightenings are shown by arrow and shaded regions, respectively. }
\label{bolo5}
\end{figure*}

Fig.~\ref{bolo5} shows the various contributions of the internal and reverse shocks to the bolometric lightcurve in the most realistic simulations: \textit{run5} (left), and in \textit{run5b} (right), which is the same run with a lower external density. As described above, several regions can lead to flare-like features: the internal forward shock, which produces the narrowest spikes, as suggested in 
\citet{2015arXiv150308333H}, and the reverse shock.
In \S\ref{sec:discussion} we will detail the observable properties of all the variability observed in our 
\textit{run5} and \textit{run5b},
and compare with the properties of the flares observed in the X-ray afterglow.

\subsection{Validity of the ballistic model}

The first goal of this study is to establish the validity of the ballistic model used by \citet{2015arXiv150308333H} to describe the dynamics of gamma-ray bursts ejecta and more specifically flares in the early afterglow phase. The bolometric light curves in Fig.~\ref{fig:lightcurves}  show a good qualitative agreement in all cases between the ballistic model and simulation. The light curves show  similar slopes, timing for flares and relative luminosity between different configurations. However, we systematically find an increased luminosity in the hydrodynamic simulations for the reverse and internal shocks, and a slight delay for the deceleration of the forward shock. The increased luminosity may be due to the uncertainty in the definition of the upstream and downstream media. However, as we have detailed in \S\ref{sec:shock_detection}, we are confident that our shock detection method yields the proper values soon after the beginning of the simulation. We have also tested that artificially increasing or decreasing $L_{\mathrm{diss}}$ and/or $\Delta t_{\mathrm{obs}}$ (see Eq.~\ref{eq:Ldiss}) at early times does not affect the luminosity at later times and cannot account for the discrepancy. As such, we expect that the difference between the ballistic model and our simulations reveals the weaknesses of the ballistic model. Specifically, the ballistic model underestimates the density in the shocked medium due to a poor spatial resolution intrinsic to the method, as already discussed in \citet{2000A&A...358.1157D}. In addition, shock reflections at late times, which strongly impact the late flare in \textit{run4} and \textit{run5} at $t_\mathrm{obs}\ge 70\, \mathrm{s}$ as seen in Fig.~\ref{bolo5} are missing in the ballistic model. Their impact was predicted by \citet{2015arXiv150308333H} (see their \S 4) and is confirmed here. 

Despite these discrepancies, the overall shapes of the curves and their respective order is the same in the model and the hydrodynamic simulations. This validates the determination of the forward shock in the ballistic model and our numerical setup with moving boundaries. The ballistic model is ideally suited to explore a wide range of parameters at negligible computational cost. However, the small quantitative differences may hinder the derivation of microphysical parameters based on direct comparison with observations. 

\section{X-ray Flares from a LLRS in a structured ejecta}\label{sec:discussion}

Based on our 1D hydrodynamic simulations,  we now explore the corresponding X-ray variability. We assume the X-ray emission results from synchrotron emission from electrons accelerated at the different shocks. Non-thermal electrons are assumed to follow a power-law distribution above a minimum Lorentz factor $\gamma_m$, with $n(\gamma)\propto \gamma^{-p}$ and $p=2.3$.
Throughout this whole section, we consider following microphysical parameters in all internal shocked regions (RS, IFS, IRS): we assume that the acceleratated electrons result from the injection of a fraction $\epsilon_\mathrm{e}=0.1$  of the dissipated energy into a fraction $\zeta=0.01$ of the electrons, and we assume that a fraction $\epsilon_\mathrm{B}=0.1$ of the  energy is injected in the amplified magnetic field; (ii) we adopt the same parameters in the external forward shock (FS), except that $\epsilon_\mathrm{e}, \epsilon_\mathrm{b}$ are both divided by $50$ to model a radiatively inefficient ultra-relativistic shock. 
As shown in Figs.~\ref{fig:lightcurves} and~\ref{bolo5}, the energy dissipated at the forward shock (FS) is higher than in the internal shocks (IFS+IRS+RS).  As such, having a radiatively less efficient forward shock is a  necessary condition for the flares to be observable in our model. As discussed in \citet{2007ApJ...665L..93U,2007MNRAS.381..732G,2012ApJ...761..147U}, this can happen if the forward shock is initially radiatively inefficient in the ultra-relativistic regime, due to an inefficient acceleration of electrons and/or an inefficient amplification of the magnetic field in the shocked external medium.

The luminosity at observed frequency $\nu_\mathrm{obs}$ is given by a modified version of Eq.~\ref{eq:lbol} \citep{1999ApJ...523..187W}:
\begin{eqnarray}
  \label{eq:lkeV}
    L_{\nu_\mathrm{obs}}(t_{\mathrm{obs}})&\!\!\!\!=\!\!\!\!&\int \mathrm{d}t  \frac{2 \epsilon_e \,  L_{\mathrm{diss}}(t) }{\Delta t_{\mathrm{obs}}(t)\, \nu_{p,\mathrm{obs}}(t,t_{\mathrm{obs}})} 
    \mathcal{B}\left(\frac{\nu_\mathrm{obs}}{\nu_{p,\mathrm{obs}}(t,t_{\mathrm{obs}})}\right)
    \nonumber\\
   & & \,\,\,\,\,\,\,\times \, \frac{1}{\left(1+\frac{t_{\mathrm{obs}}-t_{\mathrm{obs},0}(t)}{\Delta t_{\mathrm{obs}}(t)}\right)^3}\, ,
   \label{eq:Lnu}
\end{eqnarray} 
where $\nu_{p,\mathrm{obs}}$ is the peak frequency of the emission, in the observer frame,
\begin{equation}
\nu_{p,\mathrm{obs}}(t,t_{\mathrm{obs}})=\frac{2\Gamma_*(t)\, \max{\left(\nu_m(t),\nu_c(t)\right)}}{1+\frac{t_\mathrm{obs}-t_\mathrm{obs,0}(t)}{\Delta t_{\mathrm{obs}}(t)}}\, .
\end{equation}
The normalized spectral shape $\mathcal{B}(x)$ in Eq.~\ref{eq:Lnu} depends on the characteristic frequencies $\nu_m$, $\nu_c$ in the comoving frame. The latter are the synchrotron frequencies corresponding to the minimal electron Lorentz factor $\gamma_m$, and to $\gamma_c$, the Lorentz factor beyond which cooling is important, i.e.
\begin{equation}
\nu_{m,c}=\frac{3}{4\pi}  B \frac{q_e}{m_e c} \gamma_{m,c}^2,
 \end{equation}
where
\begin{eqnarray}
\gamma_m&=&\frac{\epsilon_e}{\zeta}\frac{m_p}{m_e}\frac{p-2}{p-1}\frac{\epsilon_*}{c^2}\\
\gamma_c&=&\frac{6\pi m_e c}{\sigma_T t_\mathrm{dyn} B^2},
\end{eqnarray}
with $B=\sqrt{8\pi \epsilon_b \epsilon_* \rho_*}$  the magnetic field 
and $t_\mathrm{dyn}=\frac{r}{\Gamma_* c}$ the dynamical timescale
in the comoving frame of the shocked region.
Following \citet{1998ApJ...497L..17S}, we have 
for $\nu_m\ge \nu_c$ (fast cooling):
\begin{equation}\label{eq:fast_cool}
\mathcal{B}(x)\propto \left\{ \begin{array} {ccc}\left(\frac{\nu_c}{\nu_m}\right)^{-5/6}x^{1/3}  & \textrm{ if} & x\leq \frac{\nu_c}{\nu_m}\\
x^{-1/2} & \textrm{if} & \frac{\nu_c}{\nu_m}\leq x \leq 1 \\
x^{-p/2} & \textrm{if} & x \ge 1
\end{array}
\right.\, ,
\end{equation} 
and for 
$\nu_m \le \nu_c$ (slow cooling): 
\begin{equation}\label{eq:slow_cool}
\mathcal{B}(x)\propto \left\{ \begin{array} {ccc}\left(\frac{\nu_m}{\nu_c}\right)^{-p/2+1/6}x^{1/3}  & \textrm{ if} & x\leq \frac{\nu_m}{\nu_c}\\
x^{-(p-1)/2} & \textrm{if} & \frac{\nu_m}{\nu_c}\leq x \leq 1 \\
x^{-p/2} & \textrm{if} & x \ge 1
\end{array}
\right.\, .
\end{equation} 
In both cases, the function is normalized by $\int_0^\infty \mathcal{B}(x)\, \mathrm{d}x=1$.

\begin{figure*}
\centering
\includegraphics[width=.45\textwidth]{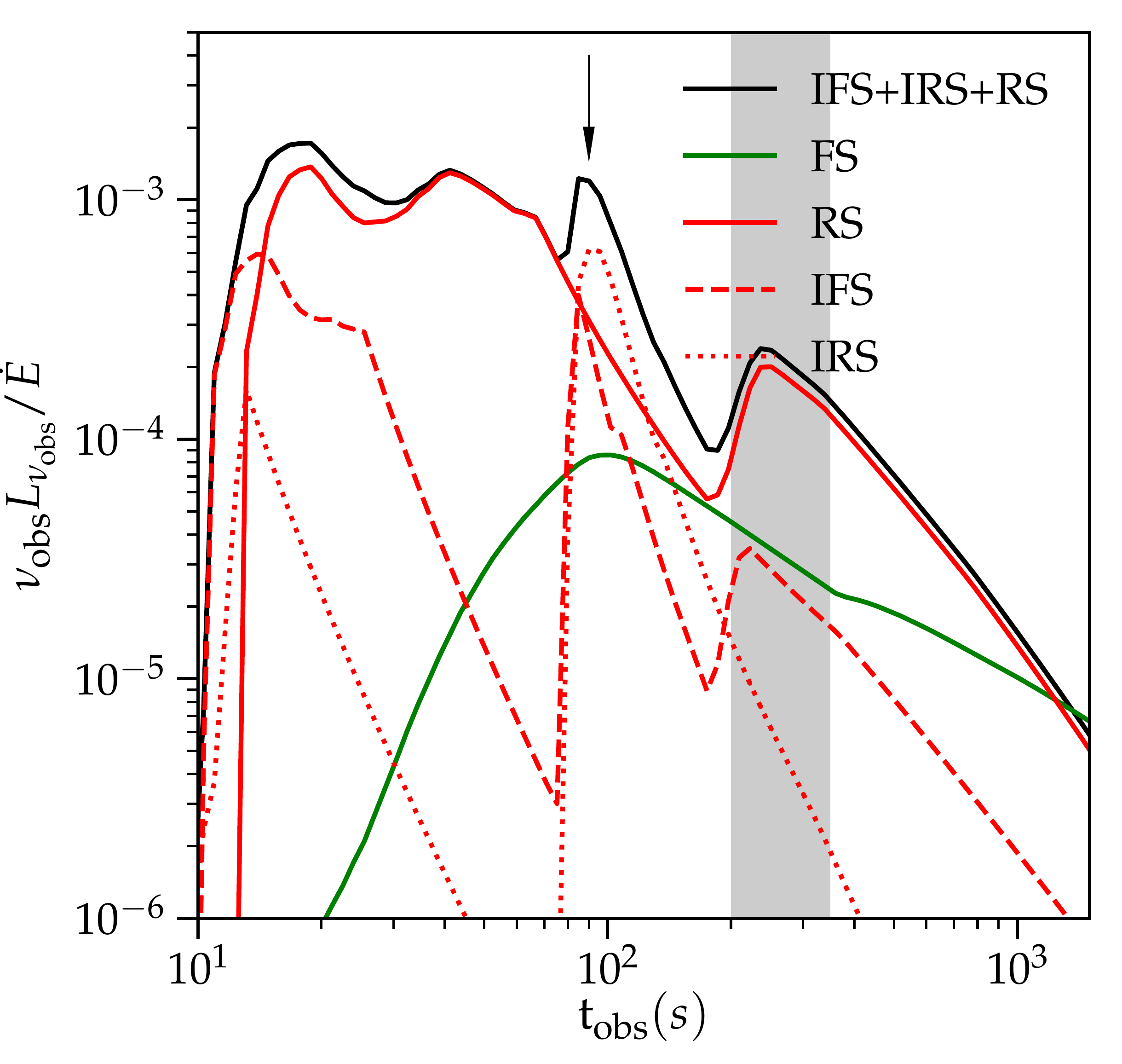}
\includegraphics[width=.45\textwidth]{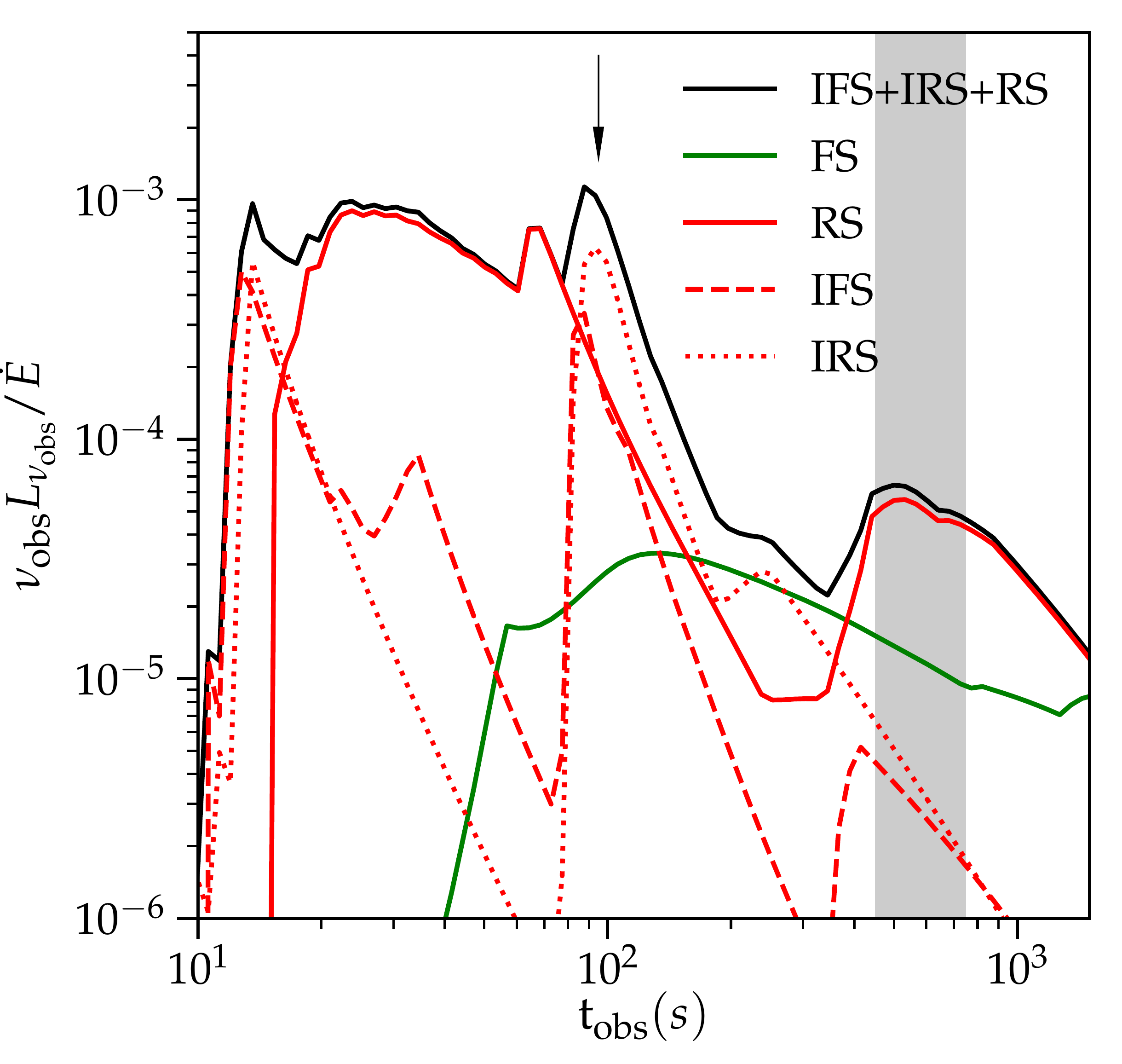}
\caption{\textbf{Separate contributions to the X-ray light curve in \textit{run5} (left) and \textit{run5b} with a lower external density (right).} The total contributions of the internal energy dissipation is shown in black. The shaded regions and arrows indicate the flares. The lightcurves are computed at $h\nu_\mathrm{obs}=1\, \mathrm{keV}$ assuming a source redshift $z=1$.}
\label{fig:compare}
\end{figure*}

Fig.~\ref{fig:compare} shows the contribution of all shocks to the lightcurve  at 1 keV in \textit{run5} and \textit{run5b}. The left plot shows our standard run, while the right panel shows a simulation with a lower external density. This figure is the X-ray  counterpart of the bolometric emission shown in Fig.~\ref{bolo5}. We clearly see three phases: (1) \textit{Prompt emission phase.}
In both cases, the very early X-ray lightcurves is dominated by internal shocks for $t_{\mathrm{obs}}\lesssim t_\mathrm{w}=100\, \mathrm{s}$; (2) \textit{Early X-ray afterglow.} For our choice of microphysics parameters, the internal energy dissipation still dominates the emission for a few thousands seconds, mainly due to the activity of the reverse shock (RS), but also with significant contributions from the internal  shocks. At this stage, the contribution of the forward shock to the emission is negligible; (3) \textit{Standard afterglow.} At later times, the external forward shock (FS) dominates. The transition occurs at $t_{\mathrm{obs}}\gtrsim 1500\,\mathrm{s}$ in \textit{run5} and appears to be rather smooth, in agreement with observations. The exact time of the transition depends of course on our arbitrary choice of microphysics parameters and could be further delayed  if ultra-relativistic shocks are strongly radiatively inefficient (i.e. for even lower values of $\epsilon_\mathrm{e}$ and/or  $\epsilon_\mathrm{B}$ in the forward shock). In addition, the transition is also delayed for a lower external density, as clearly seen in \textit{run5b}.

We now focus on the second phase, the early X-ray afterglow, to analyze the observed variability. In both \textit{run5} and \textit{run5b}, the ejecta produces two flares  shown with an arrow and shaded grey region in Fig.~\ref{fig:compare}. The first narrow spike at $t_\mathrm{obs}\simeq 90\, \mathrm{s}$ is dominated by the emission from internal shocks (IFS and IRS) before any interaction with the reverse shock, and is therefore almost identical in both simulations, in their timing as well as intensity. This confirms that late internal shocks can produce flares in the very early afterglow, especially those observed during the early steep decay phase. However, the same mechanism would require a long-lasting central engine to also produce late flares \citep{2005Sci...309.1833B,2005MNRAS.364L..42F,2006ApJ...642..354Z}.

The second, weaker, flare occurs at later times, $t_\mathrm{obs}\simeq 250\, \mathrm{s}$ in \textit{run5} and $t_\mathrm{obs}\simeq 550\, \mathrm{s}$ in \textit{run5b}, i.e. at observer times much longer than the duration of the relativistic ejection by the central engine, $t_\mathrm{w}=100\, \mathrm{s}$. For our choice of microphysics parameters in the internal regions, the emission is dominated by the reverse shock (RS). More precisely in \textit{run5} the increased emission from the reverse shock starts at $t_{\mathrm{obs}}=181$\,s, when it enters the overdense internal region (label '6' in Fig.~\ref{fig:run5}) and drops steeply after $t_{\mathrm{obs}}=335$\,s, when it exits the shocked region (label '8' in Fig.~\ref{fig:run5}). However, as mentionned above, the interaction of the reverse shock with the shocked region also causes the reflection and some rebrightening of the forward internal shock  (IFS) towards the front of the ejecta. The emission from the internal forward shock abruptly drops at $t_{\mathrm{obs}}=250$\,s, when it has reached the forward shock (label '10' in Fig.~\ref{fig:run5}). This emission always remains well below the emission of the reverse shock, but may contribute more significantly for another choice of microphysics parameters. As this second flare is due the interactions of the reverse shock with the internal dense shells, it happens later in \textit{run5b}, which has a lower external density and a later development of the reverse shock. These lightcurves confirm the scenario proposed by \citet{2015arXiv150308333H} and illustrate  the capacity of this scenario to produce late flares without invoking a long-lasting central engine.

\begin{figure*}
\centering
\includegraphics[width=.45\textwidth]{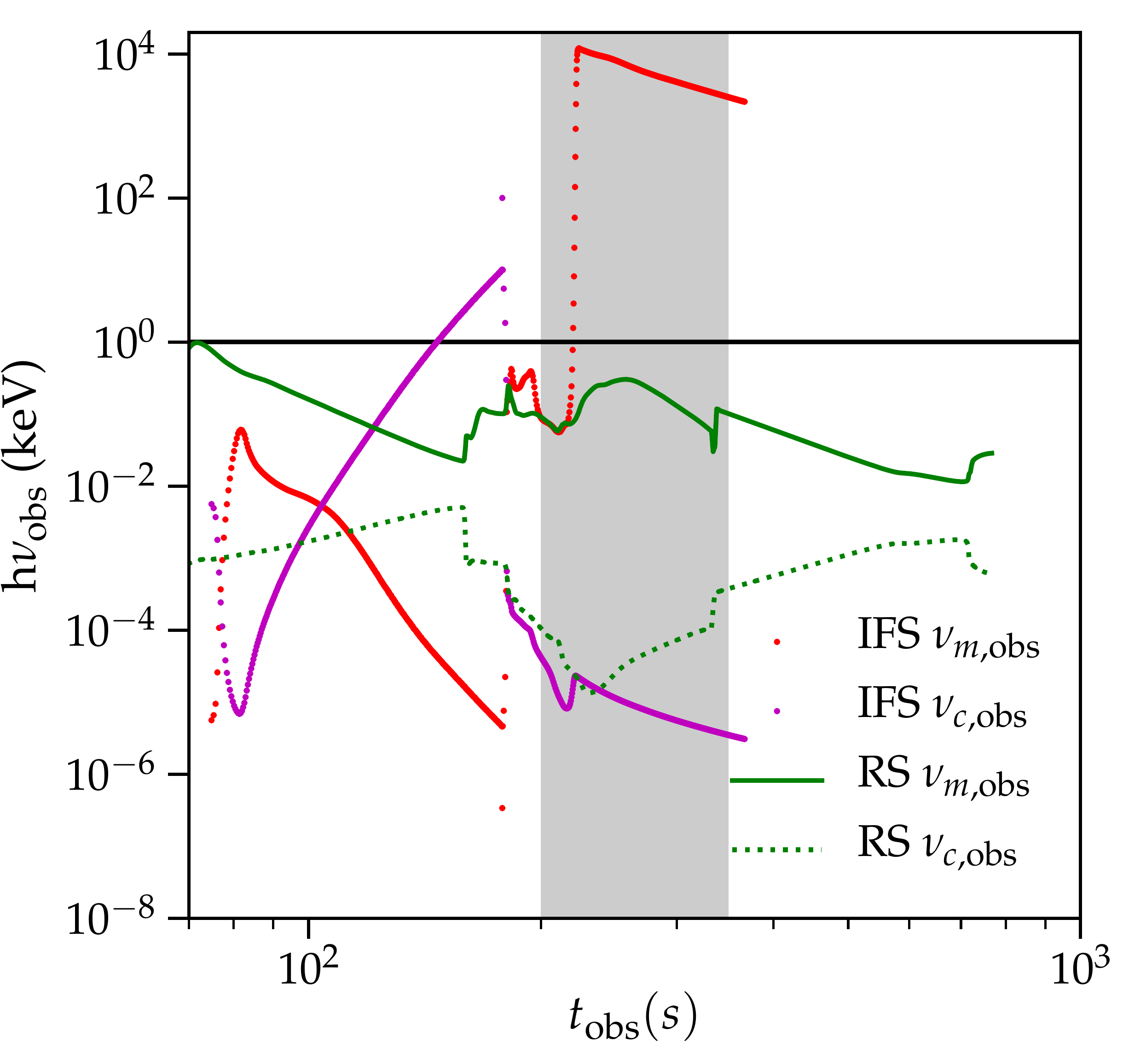}
\includegraphics[width=.45\textwidth]{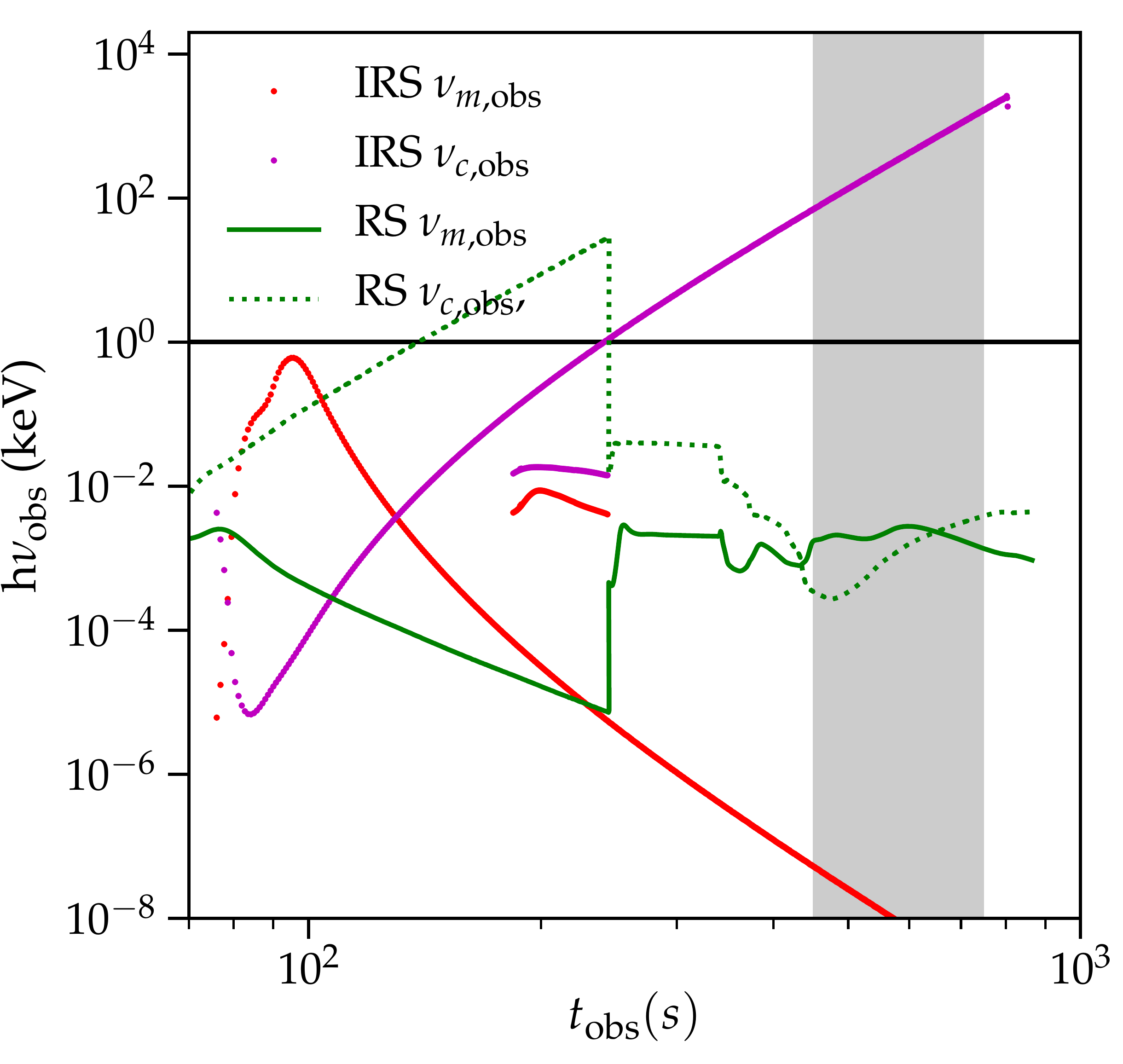}
\caption{\textbf{Characteristic synchrotron energies of the electrons at the shocks dominating the emission during the flares : RS/IFS in \textit{run5} (left)  and RS/IRS in \textit{run5b} (right).} The black line shows $h\nu_\mathrm{obs}=1\,\mathrm{keV}$ radiation for comparison. The shaded area shows the time interval of the flare.}
\label{fig:energy}
\end{figure*}

As this second flare is the novel feature of the proposed scenario, we have carefully checked our assumptions on the radiative regime. Fig.~\ref{fig:energy} shows the characteristic frequencies $\nu_\mathrm{m,obs}$ and $\nu_\mathrm{c,obs}$ from $t_\mathrm{obs}=70$ to $10^{3}\, \mathrm{s}$ and for the on-axis emission of the two dominant shocks in this period (RS/IFS in \textit{run5} and RS/IRS in \textit{run5b}). For our choice of microphysics parameters, it shows that during the second flare, the reverse shock (RS) which dominates the flare emission is in fast cooling with $h\nu_\mathrm{obs}=1\, \mathrm{keV}>\nu_\mathrm{m,obs}>\nu_\mathrm{c,obs}$. In \textit{run5b}, the RS emission enters in slow cooling during the decay phase, but electrons radiating in X-rays are still fast-cooling as $h\nu_\mathrm{obs}=1\, \mathrm{keV}>\nu_\mathrm{c,obs}$. This validates our assumptions to compute the observed lightcurve showing an X-ray flare\footnote{Fig.~\ref{fig:energy} shows that, apart from the flares, X-ray photons can be produced by slow-cooling electrons, especially at late times. In this situation, the shape of the light curve should be computed more accurately as we assume here that radiating electrons are located at the shock  (Eq.~\ref{eq:lbol}), whereas slow-cooling electrons are still radiating long after having be accelerated (see \citealt{2005ApJ...627..346B,2012ApJ...761..147U} for a more accurate method to compute the observed flux in this case). This tends to smooth the variability.} Interestingly, it appears that in \textit{run5} the IFS, the second brightest contributor, is also in fast-cooling with a peak energy in gamma-rays ($\sim 10\, \mathrm{MeV}$), which may produce a weak flare in high-energy gamma-rays. To check if the IFS could produce a gamma-ray flare such as the one detected by \textit{Fermi}/LAT in GRB 100728A \citep{2011ApJ...734L..27A,2015ApJ...803...10T}, one would require a more detailed radiative calculation including the inverse Compton scatterings.

\begin{figure}
\centering
\includegraphics[width=.45\textwidth]{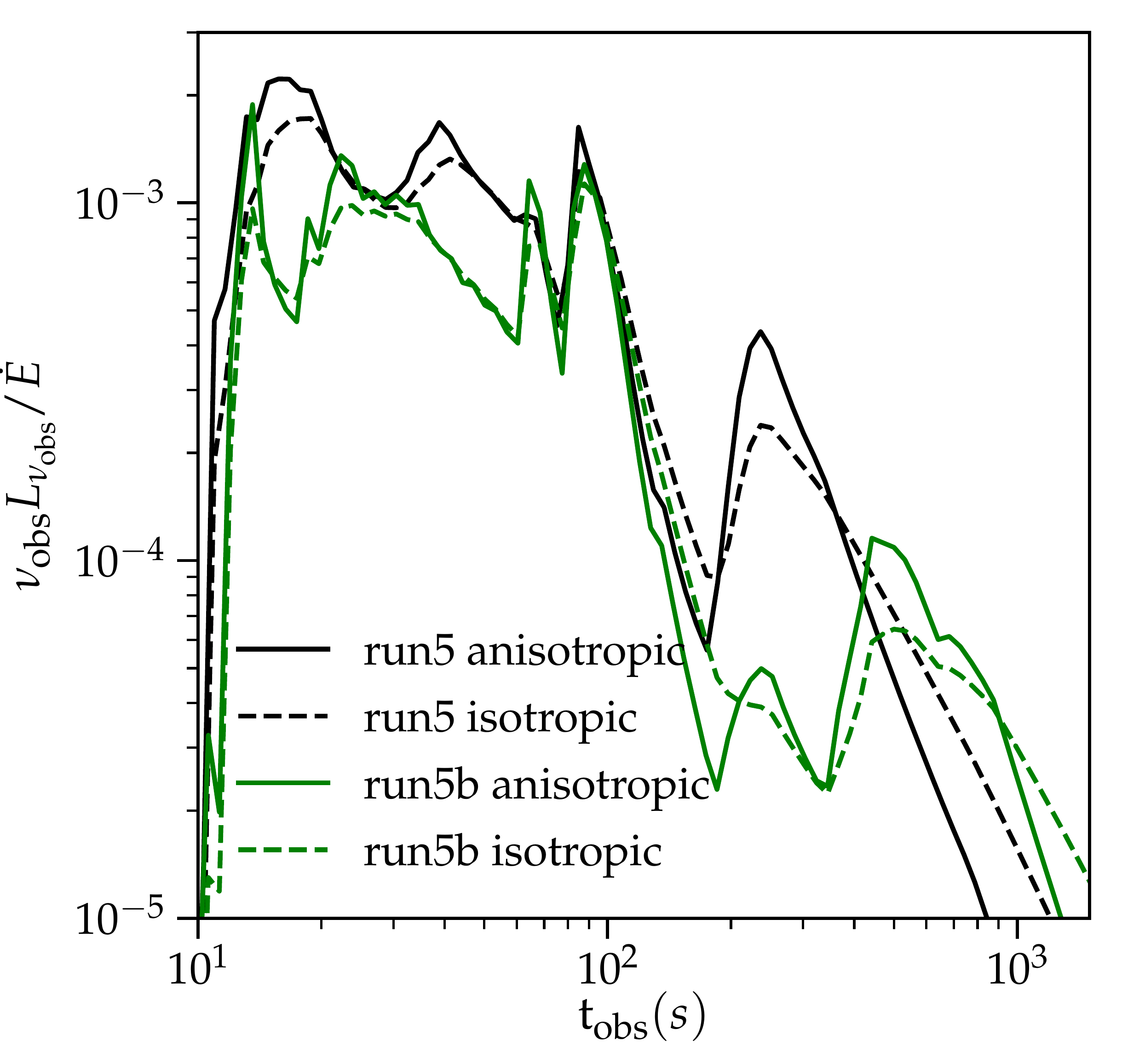}
\caption{\textbf{X-ray light curves resulting from internal energy dissipation (RS+IFS+IRS)
for \textit{run5} and \textit{run5b},
assuming an anisotropic synchrotron emission in the comoving frame: see text.} The dashed curves show the isotropic case for comparison, already shown in Fig.\ref{fig:compare}.}
\label{fig:anisotropy}
\end{figure}

 Observed X-ray flares typically have a width $\Delta t_{\mathrm{obs}}/t_{\mathrm{obs}}\simeq 0.1-0.3$ \citep{2007ApJ...671.1903C}, with a fast rise and a slower decay. However, in our simulations,  $\Delta t_{\mathrm{obs}}/t_{\mathrm{obs}}\simeq 0.7-1$. A possible solution is an anisotropic synchrotron emission in the comoving frame, as suggested by \citet{2011MNRAS.410.2422B}.  Indeed, Eq.~(\ref{eq:lbol}) clearly shows that the flare cannot decay faster than $t_\mathrm{obs}^{-3}$ in the isotropic case, whereas a steeper slope can be obtained if $A(\theta)$ decreases with $\theta$. Following \citet{2015arXiv150308333H}, we have considered the effect of a moderate anisotropy assuming $ A(\theta)\propto e^{\cos{\theta}/\cos{\theta_0}}$ with $\theta_0=70^\mathrm{\circ}$ (which leads to 90\% of the energy  being beamed within $\theta_0$). Fig.~\ref{fig:anisotropy} compares the light curves for \textit{run5} and \textit{run5b} in the anisotropic and in the standard (isotropic) case.  The latter shows more temporally peaked emission, with a higher peak luminosity resulting from limb darkening, in a much better agreement with observations. It remains to be understood if such a moderate anisotropy can be achieved in mildly relativistic shocks.

\section{Conclusion}\label{sec:conclusion}

In this paper, we present a set of high resolution one-dimensional relativistic hydrodynamic simulations of a stratified GRB ejecta interacting with a uniform external medium in order to study the variability of the early afterglow. We follow the power dissipated in each shocked region (external forward shock, internal shocks, reverse shock) and compute the resulting bolometric light curves, 
and X-ray lightcurve assuming synchrotron radiation of shock accelerated electrons in the fast-cooling regime.

The resulting bolometric light curves agree well with a ballistic model \citep{1998MNRAS.296..275D,2015arXiv150308333H}. In particular, the ballistic model is able to reproduce the slopes and relative variations of the light curves, albeit with a slightly lower total emission. However, our simulations provide a more accurate description of the dynamics: the ballistic model underestimates the density in the shocked medium, as already pointed out by \citet{2000A&A...358.1157D}, and neglects shock reflections at late times.  This validates its use for preliminary studies, as its low computing time allows for the exploration of a large range of parameters, but points out its limitation, that can impact the determination of microphysics parameters when comparing the model with observations.

We have used these simulations to validate the scenario proposed by \citet{2015arXiv150308333H} for the origin of X-ray flares in GRB afterglows. With a much more detailed description of the dynamics, our results confirm that the propagation of a long-lived reverse shock in a stratified relativistic ejecta leads to the appearance of bright flares in the bolometric light curve of the internal dissipation (i.e. all shocks except for the external forward shock). The initial stratification of the ejecta naturally results from the internal shock phase. Internal shocks locally compress the ejecta and smoothen variations of the Lorentz factor, leading to the formation of dense shells with rather uniform Lorentz factors. Early X-ray flares can be due to late internal shocks. However the most interesting feature is the appearance of late flares when the reverse shock interacts with the dense shells in the ejecta. 

Compared to the approach used by \citet{2015arXiv150308333H}, our simulations allow to accurately model this interaction, which includes shock reflections that were predicted but could not be captured by the ballistic model. Adding an estimate of the synchrotron emission, we find that the variability in the bolometric lightcurves translates into X-ray flares with properties in agreement with observations.
More specifically, the simulations  show that: 
\begin{enumerate}
\item Late internal shocks can be a source of early variability in the afterglow.
\item The interaction of the long-lived reverse shock with a dense shell in the ejecta yield a strong flare-like rebrightening,  with a fast rise and slower decay. 
\item This bright flare is not only observed in the bolometric lightcurve, but also in X-rays assuming standard microphysics parameters. Indeed, the electrons radiating at $1\,\mathrm{keV}$ are fast cooling. 
\item  Assuming a moderate anisotropy of the synchrotron emission  in the comoving frame, 
the width of this flare is of the order of $\Delta t_\mathrm{obs}/t_\mathrm{obs}\sim 0.1-0.3$, in agreement with the properties of observed X-ray flares.
\item The time at which the flare is observed depends only on the initial properties of the ejecta (distribution of the Lorentz factor, kinetic energy) and on the external density. Flares can be observed at late time, without any need for a long-lasting central engine.
\end{enumerate}

For this promising scenario to work, there is no constraint on the lifetime of the central engine as long as it is comparable to the GRB duration. On the other hand, several other assumptions are necessary:
\begin{itemize}
\item The GRB ejecta must be initially variable, to allow for the formation of internal shocks and the stratification of the ejecta with dense shells. 
\item A long-lived reverse shock must form. This is naturally expected if the central engine switches off smoothly, ejecting a tail of low-Lorentz factor material behind at the end of the relativistic ejecta. Variability of the Lorentz factor in this slow tail is required to produce the necessary stratification for shock interaction to occur on the observed timescale.  According to our model, more complex initial variability can naturally lead to multiple flares when the reverse shock crosses the different dense shells.
\item The emission of this long-lived reverse shock must dominate over the emission of the external forward shock, at least at early times when the X-rays are emitted. This is probably the strongest assumption but it is possible if  electron acceleration and/or magnetic field amplification is inefficient in the shocked external medium, behind the ultrarelativistic  forward shock.
\end{itemize}

Additional simulations are needed to explore a larger range of parameters, both for the ejecta and the external medium (such as wind-like configurations). This would allow to confirm that a large range of arrival times can be obtained for flares, as is found using the ballistic model. However, the set of simulations presented in this paper already illustrates the capacity of the long-lived reverse shock model to explain the observed diversity and variability of GRB early afterglows.

\section*{Acknowledgments}
The authors thank Zakaria Meliani for helpful advice on the setup of the simulations and Robert Mochkovitch for his insight on this manuscript. The authors thank the anonymous referee for comments that greatly improved the discussion of the X-ray properties of the ejecta. The authors thank the French Programme National Hautes Energies (PNHE) and the Centre National d'Etudes Spatiales (CNES) for financial support. Support for A. Lamberts was provided by an Alfred P. Sloan Research Fellowship, NASA ATP Grant NNX14AH35G, and NSF Collaborative Research Grant 1411920 and CAREER grant 1455342.  Numerical simulations were run on supercomputer Pleiades from the NASA Supercomputing Division and supercomputer Stampede from the Texas Advanced Supercomputing Center.

\bibliographystyle{mnras}
\bibliography{grb_1d}
\end{document}